\documentclass[a4paper,11pt]{article}
\usepackage{jcappub}
\usepackage{tikz,xcolor,hyperref}
\usepackage{amsmath, amssymb, amsthm, graphicx, epsfig, fancyhdr,epsfig, slashed}
\usepackage[normalem]{ulem}
\usepackage{tikzsymbols}
\usepackage{natbib}
\usepackage{float}
\usepackage{makecell} 

\def\m{\mu}
\def\n{\nu}						
\def\r{\rho}	
\def\s{\sigma}					
\def\levi{\epsilon}

\newcommand{\Trh}{T_\text{RH}}

\newcommand{\arh}{a_\text{RH}}

\newcommand{\aend}{a_\text{end}}
\newcommand{\rend}{\rho_\text{end}}
\newcommand{\rRH}{\rho_\text{RH}}
\def\be{\begin{equation}}
\def\noi{\noindent}
\def\ee{\end{equation}}
\def\bea{\begin{eqnarray}}
\def\eea{\end{eqnarray}}
\def\bi{\begin{itemize}}
\def\ei{\end{itemize}}
\def\beq{\begin{equation}\begin{aligned}}
\def\eeq{\end{aligned}\end{equation}}

\def\gev{\, {\rm GeV}}

\usepackage[compat=1.1.0]{tikz-feynman}
\numberwithin{equation}{section}

\begin{document}
%%%%%%%%%%%%%%%%%%%%
\title{Reheating and Leptogenesis after
Vector inflation}
%%%%%%%%%%%%%%%%%%%%%%%%%%%%%%%
\author[a]{Simon Cléry,}
\author[b]{Pascal Anastasopoulos,}
\author[a]{Yann Mambrini,}

\affiliation[a]{\,Universit\'e Paris-Saclay, CNRS/IN2P3, IJCLab, 91405 Orsay, France}
\affiliation[b]{
Institute of High Energy Physics, Austrian Academy of Sciences,\\
Georg-Coch-Platz 2, 1010 Vienna,
Austria}

\emailAdd{simon.clery@ijclab.in2p3.fr}
\emailAdd{Paschalis.Anastasopoulos@oeaw.ac.at}
\emailAdd{yann.mambrini@ijclab.in2p3.fr}

%%%%%%%%%%%%%%%%%
\abstract{We study the reheating and non-thermal leptogenesis in the case of a vector inflaton. We concentrate on particle production during the phase of oscillating background, especially gravitational production induced by the presence of non-minimal coupling 
imposed by an isotropic and homogeneous Universe. Including
processes involving the exchange of graviton, we then extend our study to decay into fermions via direct or anomalous 
couplings. 
The necessity of non-minimal gravitational coupling and the gauge nature of couplings to fermions implies a much richer phenomenology than for a scalar inflaton.}

%%%%%%%%%%%%%%%%%%
\maketitle
%%%%%%%%%%%
\section{Introduction}
\label{sec:intro}
%%%%%%%%%%%

Inflation is one of the most successful solutions to the horizon and flatness problems 
\cite{Guth:1980zm, Starobinsky:1980te, Linde:1981mu, Mukhanov:1981xt} but inflationary models are multiple. They all 
describe with great success a phenomenon of rapid expansion by introducing one or several scalar fields, the inflaton(s) 
with an almost constant energy density  during a
slow-rolling phase \cite{Olive:1989nu,Linde:1990flp,Baumann:2009ds,Martin:2013tda,mybook}. 
In the meantime, its scalar nature ensures natural
homogeneity and isotropy up to the quantum level of perturbation.
On the other hand, the first articles in the literature already stressed the importance of solving the reheating problem 
\cite{Guth:1980zm, Sato:1980yn, Guth:1982pn}. 
Indeed, at the end of the inflationary phase, the inflaton(s) enter an oscillatory phase while in the meantime 
dissipating its energy into quanta forming the primordial plasma. This phase of conversion is called reheating 
\cite{Felder:1998vq}. Adding {\it ad-hoc} new couplings
of the inflaton fields to fermions or bosons can 
be sufficient to finalize the reheating process \cite{Garcia:2021iag,Garcia:2020wiy,Garcia:2020eof}, even at a 
gravitational level \cite{Mambrini:2021zpp,Clery:2021bwz,Haque:2022kez,Co:2022bgh} even if non-minimal couplings to the Ricci scalar 
are necessary to avoid the overproduction of primordial gravitational waves 
\cite{Clery:2022wib,Barman:2022qgt,Chakraborty:2023ocr,Barman:2023ktz}.

It is interesting to note that higher-spin bosonic fields 
have been overlooked since they naturally generate {\it a priori} inhomogeneities due to their spatial dependence. 
Whereas a first attempt at a vector inflaton was proposed in \cite{Ford:1989me}, the authors of \cite{Golovnev:2008cf} 
showed that the presence of three orthogonal vector fields ensures {\it at the same time} a slow-roll regime equivalent 
to a scalar inflaton while generating a homogeneous Universe. The price to pay is easy to guess: the need for a 
conformal coupling to the Ricci scalar to cancel the vectorial nature of the field, while the presence of 3 orthogonal 
fields ensures that no specific direction in the expansion is privileged. It becomes then interesting to look into the 
details of reheating in this framework. 

Indeed, the presence of non-minimal coupling, {\it necessary}
for the inflaton to slow-roll, ensure by default a portal between the inflaton and the standard model bath. Even if 
weak, it can complete the reheating process as it was shown for the case of a scalar inflaton in \cite{Clery:2022wib}.
On the other hand, it is also natural to imagine the presence of couplings between the vector inflatons $A_i$
and particles charged under it, especially when embedding this model into a Grand Unified (GUT) framework.
This renders the model much more natural than in the case of a scalar inflaton, where the reheating is usually 
considered through an unconstrained Yukawa-type coupling. Other decay processes are also present in 
the case of a vector inflaton and absent in the scalar case. First, even if SM fermions are not directly coupled to the vector inflaton fields, inevitable kinetic mixing induces a decay of the vector fields towards SM fermions. In addition, Chern-Simons couplings are generated by the 
decoupling of anomalies \cite{Anastasopoulos:2006cz, Anastasopoulos:2007qm, Anastasopoulos:2008jt, 
Anastasopoulos:2022ywj}. Indeed, if a gauge structure is hidden under the presence of a vector inflaton field, it is 
natural to suppose the existence of heavy fermions. If their mass lies above the inflationary scale, at the GUT scale, 
for instance, their
decoupling will generate effective three-vectorial vertices through triangle anomalies which
could generate decay or new scattering processes inducing an effective reheating.
Finally, an even more suggestive setup, especially in an SO(10) framework, would be the presence of a direct coupling 
between the inflaton fields $A_i$ and the right-handed neutrinos $N_R$. In this case, the decay process can even generate a sufficient lepton asymmetry to ensure successful baryogenesis.

The paper is organized as follows. After reminding the physics in the presence of a vector inflaton in section 2, we 
compute and analyze the 
reheating phase in section 3. We first consider purely gravitational couplings, including the exchange of graviton, then 
direct gauge couplings to fermions before looking into the details of loop-induced coupling generated by triangle 
anomalies. Finally, we look into the details of non-thermal leptogenesis, adding the presence of right-handed neutrino in section 4, 
before concluding.

%%%%%%%%%%%
\section{The framework}
\label{sec:framework}
%%%%%%%%%%%
We consider the framework of Vector inflation introduced in \cite{Golovnev:2008cf}, where cosmic inflation is driven by 
vector fields. The action in the Jordan frame is written 
\be
    S = \int d^4x \sqrt{-g}\left( -\frac{M_P^2}{2}\Omega^2 \mathcal{R}  -\frac{1}{4}F_{\mu \nu}F^{\mu \nu} + \frac{1}{2} 
    M^2 A_\mu A^\mu\right)
    \label{Eq:action}
\ee
with
\be
\label{Eq.conformal_factor}
    \Omega^2 = 1 + \frac{\xi A_\mu A^\mu}{M_P^2}
\ee
\be
    F_{\mu\nu} = \nabla_\mu A_\nu - \nabla_\nu A_\mu = \partial_\mu A_\nu - \partial_\nu A_\mu
\ee
where we defined $\mathcal{R}$ as the Ricci scalar curvature, $M$ the mass of the vector field $A_\mu$ and $\xi$ the non-
minimal coupling to gravity of this vector field \footnote{$M_P = (8 \pi G_N)^{-1/2} \simeq 2.4 \times 10^{18}$ GeV is 
the reduced Planck mass.}.
Note that $\nabla_\mu$ is the covariant derivative, 
\beq
\nabla_\mu A_\nu = \partial_\mu A_\nu -\Gamma_{\mu \nu}^\alpha A_\alpha
\nonumber
\eeq
\noi
Such models may lead to instabilities in the quasi De Sitter background, at the 
perturbation level \cite{Himmetoglu:2008zp, 
Golovnev:2008hv, Himmetoglu:2008hx, Himmetoglu:2009qi, 
Golovnev:2009rm}. Some constructions have been proposed to cure this instability through different 
modifications of potential and kinetic terms of the 
vector fields \cite{Esposito-Farese:2009wbc, Golovnev:2011yc}. However, it has also been proposed that more general modifications of gravity involving $\beta A_\mu A_\nu R^{\mu \nu}$ term as well as a coupling of the vector fields with the Gauss-Bonnet invariant $\alpha A^2\mathcal{G}$, can lead to successful inflation for different values of the couplings $(\xi, \alpha, \beta)$ \cite{Oliveros:2016myr, Bertolami:2015wir}. It is possible that the additional dynamics generated by these terms could avoid the appearance of ghost modes for longitudinal perturbations. In particular, the additional Gauss-Bonnet term can be interpreted as an additional positive contribution to vector effective mass during De
Sitter inflation through, 
\beq
\frac{1}{2}\left( M^2 + \xi \mathcal{R} + \alpha \mathcal{G} \right)
\eeq
with $\mathcal{G} \sim R^2$. This allows for suppression of the instability generated by the negative contribution from non-minimal coupling to curvature with $\xi<0$ \cite{Oliveros:2016myr}. At the end of inflation, such dimensionful coupling $\alpha$, with higher power of the curvature from the Gauss-Bonnet invariant, is expected to be suppressed in comparison with the dimensionless non-minimal coupling to curvature $\xi$, as the vector fields become sub-planckian during reheating. Nevertheless, studying the stability of the model during inflation requires full gravitational and field theory perturbation analysis
which is beyond the scope of this work. The ghost instability problem will not be considered in this work.\\

We consider the usual FLRW metric with the following signature
\be
ds^2 = dt^2 -a^2(t)\delta_{ik}dx^idx^k
\ee
which provides the following Jacobian determinant $\sqrt{-g}=a^3(t)$.
The variation of the action with respect to $A^\mu$, assuming homogeneity of the fields, i.e $\forall \alpha, \forall i, 
\partial_i A_\alpha = 0$, yields the following equations for time and spatial components of $A^\mu$
\be
\frac{1}{\sqrt{-g}} \frac{\partial}{\partial x^\mu}(\sqrt{-g}F^{\mu\nu}) + \left(M^2 -\xi\mathcal{R} \right)A^\n=0\,,
\ee

\noi 
which solutions are
\bea
    && A_0 = 0 \\
    &&\ddot{A_j} + H\dot{A_j} + \left(M^2 -\xi\mathcal{R} \right)A_j=0
\label{Eq:eomai}
\eea
\noi
with $H=\frac{\dot a}{a}$ and 
${\cal R}=-6 \frac{\ddot a}{a}-6 H^2=-6 \dot H -12 H^2$. From 
\beq
A^\alpha A_\alpha = A_0^2-\frac{A_i^2}{a^2}=-\frac{A_i^2}{a^2}\,,
\nonumber
\eeq
it is natural to redefine a normalized conformal vector field $B_\mu=\frac{A_\mu}{a}$, which does not "feel" the effect 
of the expansion under the Lorentz transformation $B^\mu = g^{\mu \nu}B_\nu$. It should behave as a scalar field in the 
conformal theory. Indeed, the equation 
(\ref{Eq:eomai}) then becomes
\beq
\ddot B_i + 3 H \dot B_i + M^2 B_i +(1+6 \xi)
\left[\dot H + 2H^2\right]B_i=0\,.
\label{Eq:eombi}
\eeq
In the same way, we can also extract from the action (\ref{Eq:action})
the energy density of the field $B_i$, $T_{00}$ with
\beq
T_{\mu \nu} = \frac{2}{\sqrt{-g}}\frac{\delta {\cal S}}{\delta g^{\mu \nu}}\,.
\eeq
We obtained
\beq
T_{00}=\frac{1}{2} \dot B_i^2 + \frac{M^2}{2}B_i^2 
+\frac{H^2}{2}(1+6 \xi) B_i^2+(1+6\xi)H B_i \dot B_i \,.
\label{Eq:t00bi}
\eeq

The interesting point concerning the above equations (\ref{Eq:eombi}) and (\ref{Eq:t00bi}), 
is that whereas the non-minimal coupling $\xi=-\frac{1}{6}$ converts
a massless scalar field into a conformal invariant field, the effect here is the opposite. It {\it violates} the 
classical conformal invariance of a massless vector field.
Hence if we set $\xi =  -\frac{1}{6}$, we recover the same equations of motion and the energy density for $B_i$ as for a minimally coupled scalar inflaton. 
In other words, a non-minimally coupled vector field with $\xi =  -\frac{1}{6}$ leads to the same dynamics as a 
minimally coupled scalar field in the cosmological background. 
Hence, it can mimic the inflaton dynamics in the standard framework of slowly rolling single scalar field $\phi$, in a 
potential $V(\phi)$, which exhibits a plateau for large field values, with $\phi=B_i$. 
However, to avoid a preferred direction for inflation which the vector field direction would give, one needs to 
introduce at least three mutually orthogonal vector fields $B_i^{(a)}$, $a=1,2,3$, as it has been noted in 
\cite{Golovnev:2008cf}.\\

To understand this specificity, we have to compute the pressure 
from the stress-energy tensor $T_{ij}$. We obtained
\bea
&&
\frac{1}{a^2}T_{ij} =
\left(M^2+6 \xi \dot H + (12 \xi-1)H^2\right)
B_iB_j-\dot B_i \dot B_j
-H B_i \dot B_j - H \dot B_i B_j
\label{Eq:tij}
\\
&&
+\left[\left((2 \xi-\frac{1}{2})M^2+(9 \xi + \frac{5}{2})H^2
+12 \xi^2 \dot H\right) B_i^2
+(\frac{1}{2}-2 \xi)\dot B_i^2+(1+2 \xi)B_i\dot B_i\right]\delta_{ij}\,,
\nonumber
\eea
with the pressure $P_{ij}=T_{ij}$.
To visualize the phenomena, it is easier to choose the system of coordinates with the third axes aligned with $B_i$. In 
this
system, $B_i$ can be written as $B_i=B \delta_{iz}$. During inflation, where we can neglect the time derivative of the 
fields
and the Hubble rate, Eq.(\ref{Eq:tij}) gives for the pressure in the $x,y$ and $z$ directions respectively, for $\xi=-
\frac{1}{6}$:
\beq
P_{xx}=P_{yy} \simeq + a^2 B^2 H^2\,~~~P_{zz} \simeq -2 a^2 B^2 H^2\,,
\eeq
where we supposed $H\gg M$ during inflation. We clearly understand that such a Universe would extend exponentially in 
the $z$-direction, whereas it would be contracted in the $x$ and
$y$ directions, more like a cigar shape than an isotropic shape.

To circumvent this problem, the solution is obvious: one should symmetrize the system in all directions, adding two 
new 
orthogonal fields with the same amplitude $B$. 
It is then easy to show that the stress
tensor $T_{ij}$ (\ref{Eq:tij}) becomes
\beq
\frac{1}{a^2} T_{ii}=(\frac{1}{2}-6 \xi)\dot B^2
+\left[
(6 \xi - \frac{1}{2})M^2+(39 \xi + \frac{13}{2})H^2
+(36 \xi^2 + 6 \xi)\dot H
\right]B^2
+(1+6 \xi)H B \dot B\,.
\eeq
whereas the density of energy becomes
\beq
T_{00}=\frac{3}{2}\dot B^2 + \frac{3}{2}M^2 B^2 
+\frac{3}{2}(1+6 \xi)H \dot B +3(1+6 \xi)H B_i \dot B_i\,.
\eeq
For $\xi=-\frac {1}{6}$ one obtains
\bea
&&
\rho=\frac{3}{2} \dot B^2 + \frac{3}{2} M^2 B^2\,,
\nonumber
\\
&&
P=\frac{3}{2}\dot B^2-\frac{3}{2}M^2B^2\,.
\nonumber
\eea
It corresponds to the classical pressure and density of 
energy for a set of 3 scalar inflaton background fields, respecting the classical equation of motion (\ref{Eq:eombi}):
\beq
\ddot B + 3 H \dot B + M^2 B =0 
\eeq

To achieve an inflationary mechanism, we now introduce a specific potential for the background fields, the 
same 
for each copy of these vector fields. Of course, many possible potentials $V(|B|)$ can account for inflation. 
However, the relevant calculations during the reheating era are largely independent of the potential during inflation 
and depend only on the shape of the potential around the minimum. 
Without loss of generality, we will assume that $V(|B|)$ is among the class of $\alpha$-attractor
\be
\label{Eq.full_potential}
V(|B|) = \lambda M_P^4\left[\sqrt{6} \tanh{\left(\frac{|B|}{\sqrt{6}M_P}\right)} \right]^k
\ee
where $|B|^2=\delta_{ij}\delta_{ab}B^{(a)}_iB_j^{(b)}$. The overall scale of the potential parameterized by the coupling 
$\lambda$ can be determined from the amplitude of the CMB power spectrum $A_S$,  
\be
\label{Eq.potnormalization}
\lambda \simeq \frac{18\pi^2 A_S}{6^{k/2}N^2} ,
\ee
where $N*$ is the number of e-folds measured from the end of inflation to the time when the pivot scale $k_*=0.05~{\rm 
Mpc}^{-1}$ exits the horizon. 
In our analysis, we use $\ln(10^{10}A_S)=3.044$ \cite{Planck:2018jri} and set $N*=55$. 
This potential can be expanded near its minimum \footnote{Our discussion is general and not limited to T-models of 
inflation as the way we express the minimum of the potential is generic.} by
\be
    \label{Eq:potmin}
    V(|B|)= \lambda \frac{|B|^{k}}{M_P^{k-4}}\,; \quad |B| \ll M_P \, .
\ee
In this class of models, inflation occurs at large field values ($|B| \gg M_P$). 
After the exponential expansion period, the fields oscillate about the minimum, and the reheating process begins. 
The end of inflation may be defined when $\ddot a=0$ where $a$ is the cosmological scale factor, and we denote the 
inflaton energy density at $a_{\rm end}$ by  $\rho_{\rm end}$. \\

In addition to the inflationary sector, we first assume no other couplings besides the non-minimal coupling to gravity 
of the vector fields. 
To extract the coupling between $B$ and the Standard Model particles, we consider the conformal transformation between 
the Jordan frame and the Einstein frame, which involves the non-minimal coupling $\xi$,
\be
g_{\mu\nu}^{(E)}= \Omega^2 g_{\mu\nu}^{(J)}
\ee
where the superscripts stand respectively for the Einstein and the Jordan frame. It can be shown \cite{fujii_maeda_2003} 
that in the Jordan frame, the Ricci curvature expressed with the Einstein frame variables is given by 
\be
{\cal R}^{(J)} = \Omega^2\left[ {\cal R}^{(E)} + 6 g^{\mu\nu}\nabla_\mu\nabla_\nu\log(\Omega) - 6g^{\mu\nu}
(\nabla_\mu\log(\Omega))(\nabla_\nu\log(\Omega))\right].
\ee
Noting that $\sqrt{-g^{(J)}} = \frac{1}{\Omega^4}\sqrt{-g^{(E)}}$, the first term provides the usual Einstein-Hilbert 
action and hence usual gravity. 
The second term is a total derivative that will play no role in the action, and the last one can be expressed as a 
modification of the kinetic term for the vector fields $A^\mu$. 
In the Einstein frame, the total action that includes the Standard Model (SM) fields and additional vector fields can be expressed as follows:
\bea
\nonumber
&&S^{(E)} =\int d^4x\sqrt{-g}  \left[-\frac{M_P^2}{2}{\cal R}^{(E)} - \frac{3}{2\Omega^2M_P^4}\nabla_\mu(\xi A_\nu 
A^\nu)\nabla^\mu(\xi A_\lambda A^\lambda) -\frac{1}{4}F_{\mu\nu}F^{\mu\nu} + \frac{1}{\Omega^2}V(A_\mu A^\mu) \right. 
\\
\label{Eq.Einstein_action}
&& ~~~~~~~~~~~~~~~~~~~~~~~~~~\left. -\frac{1}{\Omega^4}V_h({\bf H}) + \frac{1}{\Omega^2}(D_\mu {\bf H})^\dag(D^\mu {\bf 
H}) + ... \right]
\eea
where ${\bf H}$ denotes the Higgs complex scalar doublet in the SM. Note that the kinetic terms are not canonical. In 
what follows, we will be interested in the small-field limit corresponding to the post-inflationary phase
\be
    \label{eq:smallfield}
    \frac{|\xi A_\mu A^\mu|}{M_P^2} \; \ll 1 \, .
\ee
In that case, we can expand the kinetic and potential terms in the action in powers of $M_P^{-2}$. We obtain a 
canonical kinetic term for the scalar and vector fields and deduce the leading-order interactions induced by the non-
minimal couplings.
The latter can be brought to the form 
\be
    \label{lag4point}
   \mathcal{L}_{\rm{non-min.}} \; = \; -\frac{\xi}{2M_P^2}\partial_\alpha h \partial^\alpha h A_\mu A^\mu \, + 
   m_h^2\frac{\xi}{M_P^2}h^2 A_\mu A^\mu \, , 
\ee
Here we neglected the quartic coupling of the Higgs field in comparison with its mass $m_h$, and we considered $N_h=4$ 
real scalar degrees of freedom $h$, embedded in the complex scalar doublet ${\bf H}$. \\

In addition to this interaction generated in the Einstein frame by the non-minimal couplings to the gravity of the 
background fields, we should also consider the unavoidable graviton exchange to produce matter and radiation 
\cite{Clery:2021bwz, Clery:2022wib, Barman:2022tzk}. In the Einstein frame, the metric 
can be expanded locally around Minkowski space-time, $g_{\mu \nu}\simeq \eta_{\mu \nu}+\frac{2h_{\mu \nu}}{M_P}$ 
\footnote{Note that this is the canonical normalization of spin-2 perturbation of the metric.}. Then the minimal 
gravitational interactions are described by the Lagrangian~\cite{Choi:1994ax,Holstein:2006bh}
\beq
\sqrt{-g}\,\mathcal{L}_{\rm min}= -\frac{1}{M_P}\,h_{\mu \nu}\,\left(T^{\mu \nu}_{\rm SM}+T^{\mu \nu}_A  \right) \,,
\label{Eq:lagrangian}
\eeq
where $A$ refers to any background vector field.
The canonical graviton propagator for momentum $p$ is
\be
 \Pi^{\mu\nu\rho\sigma}(p) = \frac{\eta^{\rho\nu}\eta^{\sigma\mu} + 
\eta^{\rho\mu}\eta^{\sigma\nu} - \eta^{\rho\sigma}\eta^{\mu\nu} }{2p^2} \, .
\ee
The form of the stress-energy tensor $T^{\mu \nu}_s$ depends on the spin $s$ of the field and, for massive vector field, takes the form
\be
\label{eq:tmunu}
T_{1}^{\mu \nu} = -\frac{1}{2} \left( F^\mu_\alpha F^{\nu \alpha} + F^\nu_\alpha F^{\mu \alpha} \right) + \frac{1}{4} g^{\mu \nu} 
F^{\alpha \beta} F_{\alpha \beta} -\frac{1}{2}g^{\mu \nu} M^2 A_\alpha A^\alpha + M^2 A^\mu A^\nu\,,
\ee
whereas for a scalar $S$,
\be
T^{\mu \nu}_{0} =
\partial^\mu S \partial^\nu S-
g^{\mu \nu}
\left[
\frac{1}{2}\partial^\alpha S\,\partial_\alpha S-V(S)\right]\,.
\label{eq:tmunuphi}
\ee

%%%%%%%%%%%%%%%%%%%%%%%%%
\section{Reheating}
\subsection{Background scattering through gravitational portals}
\label{sec:reheating_grav}
%%%%%%%%%%%%%%%%%%%%%%%%%%%%%

From the action introduced in the precedent section, we can consider the couplings of the background vector fields 
responsible for inflation, which allow to produce relativistic quanta after inflation and to reheat the Universe. We 
first compute the equivalent reheating temperature generated by the gravitational portals, including the minimal and 
non-minimal coupling to the gravity of the background vector fields that oscillate after inflation. These result in 
direct couplings to the SM Higgs fields ${\bf H}$ that would constitute the primordial plasma. The minimal process of 
graviton exchange interferes with the direct production of Higgs bosons involving non-minimal coupling to gravity
\bea
&&\scalebox{1}{
\begin{tikzpicture}[baseline={-0.1cm}]
  \begin{feynman}[every blob={/tikz/fill=gray!30,/tikz/inner sep=2pt}]
    \vertex (i1) at (-1.75, 1.2) {\(A_\m\)};
    \vertex (i2) at (-1.75,-1.2) {\(A_\n\)};
    \vertex (l1) at (-0.5, 0);
    \vertex (r1) at (0.5, 0);
    \vertex (f1) at (1.75, 1.2) {\(h\)};
    \vertex (f2) at (1.75,-1.2) {\(h\)};
    \diagram* {
      (i1) -- [boson, style=red] (l1),
      (i2) -- [boson, style=red] (l1),
      (f1) -- [scalar] (r1),
      (f2) -- [scalar] (r1),
      (r1) -- [gluon, edge label'=\(h_{\m\n}\)] (l1)}; \end{feynman}
\end{tikzpicture}}
~~~ + ~~~
\scalebox{1}{
\begin{tikzpicture}[baseline={-0.1cm}]
  \begin{feynman}[every blob={/tikz/fill=gray!30,/tikz/inner sep=2pt}]
    \vertex (i1) at (-1.75, 1.2) {\(A_\m\)};
    \vertex (i2) at (-1.75,-1.2) {\(A_\n\)};
    \vertex (m) at (0,0) {\(\xi\)};
    \vertex (f1) at (1.75, 1.2) {\(h\)};
    \vertex (f2) at (1.75,-1.2) {\(h\)};
    \diagram* {
      (i1) -- [boson, style=red] (m),
      (i2) -- [boson, style=red] (m),
      (f1) -- [scalar] (m),
      (f2) -- [scalar] (m)}; \end{feynman}
\end{tikzpicture}}
    \label{fig:AA->hh}
\eea
First, we make this observation on the background vector fields
\be
|A_\mu A^\mu|=|-\frac{1}{a^2}A_iA_i|=|B_iB_i|
\ee
as $A_0=0$ from the homogeneity constraints. Hence, in every process involving the vector fields scattering, this is 
the physical vector field $B_i$, which is involved and which has only spatial components. We separate the slow-varying 
envelop and the fast oscillating part of each vector field and then perform the Fourier 
expansion of the fast oscillating function
\be
B_i^{(a)}(t) =B(t) \,\epsilon^{(a)}_{i} \sum\limits_{n=1}^{\infty} \mathcal{P}_n e^{-in\omega t}
\label{eq:fourier}
\ee
The frequency of the oscillations can be obtained \cite{Garcia:2020wiy}
\be
\omega = m_B \sqrt{\frac{\pi k}{2(k-1)}}\frac{\Gamma(\frac{1}{k}+\frac{1}{2})}{\Gamma(\frac{1}{k})}
\ee
where we defined the time-dependent effective mass of the condensate as 
\be
m_B^2(t)=V^{''}(|B|(t))=k(k-1)\lambda M_P^2\left(\frac{|B|}{M_P}\right)^{k-2}
\ee
We consider that each vector field, $B_i^{(a)}$, $a=1,2,3$, carries only one polarization, mutually orthogonal between the three vector fields 

\be
    \epsilon_i^{(1)} = (1,0,0), ~~~~ \epsilon_i^{(2)} = (0,1,0), ~~~~ \epsilon_i^{(3)} = (0,0,1)
\ee
This is referred to as the cosmic triad formalism \cite{Armendariz-Picon:2004say}.
%\be
%\sum\limits_{\lambda=1}^3 \epsilon_{i,\lambda}\epsilon_{j,\lambda}^\ast =\delta_{ij}
%\ee
The scattering amplitude related to the production rate of the processes $A_\mu^{(a)} + A_\mu^{(a)} \rightarrow h_{\mu\nu} 
\rightarrow {\rm{SM}}^i+{\rm{SM}}^i$, can be parametrized by
\begin{equation}
\mathcal{M}^{1i} \propto M_{\mu \nu}^1 \Pi^{\mu \nu \rho \sigma} M_{\rho \sigma}^i \;, 
\end{equation}
where $i$ denotes the spin of the final state involved in the scattering process.
The partial amplitudes, $M_{\mu \nu}^i$, for two fields of momenta $(p_1, p_2)$ and polarization vectors $(\epsilon_1, \epsilon_2)$ in initial or final state, are given 
by \cite{Clery:2021bwz}
\bea 
M_{\mu \nu}^{0} &=& \frac{1}{2}\left[p_{1\mu} p_{2\nu} + p_{1\nu} p_{2\mu} - \eta_{\mu \nu}p_1\cdot p_2 - \eta_{\mu 
\nu} V''(S)\right] \,, \\ 
M_{\mu \nu}^{1/2} &=&  \frac{1}{4} {\bar v}(p_2) \left[ \gamma_\mu (p_1-p_2)_\nu + \gamma_\nu (p_1-p_2)_\mu \right] 
u(p_1) \, , \\
M_{\mu \nu}^{1} &=& \frac{1}{2} \bigg[ \epsilon_{2} \cdot \epsilon_{1}\left(p_{1 \mu} p_{2 \nu}+p_{1 \nu} p_{2 
\mu}\right)
\nonumber\\
&&-\epsilon_{2} \cdot p_{1}\left(p_{2 \mu} \epsilon_{1 \nu}+\epsilon_{1 \mu} p_{2 \nu}\right) - \epsilon_{1} \cdot 
p_{2}\left(p_{1 \nu} \epsilon_{2 \mu}+p_{1 \mu} \epsilon_{2 \nu}\right)
\nonumber\\
&&+\left(p_{1} \cdot p_{2} +V''(B)\right)\left(\epsilon_{1 \mu} \epsilon_{2 \nu}+\epsilon_{1 \nu} \epsilon_{2 
\mu}\right)   \nonumber \\
&&+\eta_{\mu \nu}\left(\epsilon_{2} \cdot p_{1} \epsilon_{1} \cdot p_{2}-\left(p_{1} \cdot p_{2} + V''(B) \right)\, 
\epsilon_{2}\cdot \epsilon_{1}\right) \bigg]  \, ,
\label{partamp}
\eea
We obtain the transition amplitude involving the scattering of the background oscillating modes through the gravitational portals, for the contribution of one of the three vector fields labeled $(a)$,
\be
-i\mathcal{M}_n^{(a)} \simeq -i \frac{|B|^2}{M_P^2}\mathcal{P}^{(2)}_n\left( M_{\mu\nu}^1 \Pi^{\mu\nu\rho\sigma} 
M_{\rho\sigma}^0 + \xi \frac{E_n^2}{4} \right)
\ee
where we have neglected the Higgs mass $m_h$ in comparison with energy available in the vector condensate 
$\sqrt{s}=E_n=2n\omega \gg m_h$ \footnote{Here $s$ is the Mandelstam variable.}. We have also introduced the Fourier coefficients, $\mathcal{P}_n^{(2)}$, associated with the expansion of the function $B(t)^2$, in opposition with $\mathcal{P}_n$, that are associated with the Fourier expansion of $B(t)$ as defined in \ref{eq:fourier}. \\

Then, summing over identical final states, 
symmetrizing, we recover the square amplitude \footnote{The factors of 2 in front of the amplitude square accounts for the sum over identical final states and symmetry.}
\be
|\bar{\mathcal{M}}_n^{(a)}|^2 = \frac{2^2}{2}\frac{|B|^4}{64 M_P^4}|\mathcal{P}^{(2)}_n|^2 \left(E_n^4(2\xi-1)^2 +36(\epsilon_i^{(a)}\,p_i)^4 -12E_n^2(\epsilon_i^{(a)}\,p_i)^2(2\xi-1) \right)
\ee
where $p_i$ are the spatial components of the outgoing momentum. Integrating the phase space with the initial state at rest and the two-body final state, we have the following rate 
of particle production
\be
R = 2\times 3\times \frac{
N_h}{5120\pi}\frac{|B|^4}{M_P^4}\left(80\xi^2 -120\xi +49\right)\sum\limits_{n=1}^{\infty}E_n^4|\mathcal{P}^{(2)}_n|^2
\label{eq:grav_rate}
\ee
From the amplitude of the vector fields, we can define the total energy densities they are carrying as a homogeneous 
condensate \cite{Golovnev:2008cf}
\be
\rho_{B} = \frac{3}{2}\left( \dot{|B|}^2 + 2V(|B|) \right)
\ee
where the factor of 3 accounts for the number of vector fields needed to impose a homogeneous and isotropic 
inflationary phase. 
During the reheating phase, we have to solve the following
set of coupled Boltzmann equations 
\bea
    &&\frac{d\rho_{B}}{dt} + 3(1+w_B)H\rho_B \simeq 0 \\
    &&\frac{d\rho_R}{dt} + 4H\rho_R = R\times E_n
    \label{Eq:boltzmann}
\eea
where we neglected the depletion rate of the condensate until the reheating ended in the first equation, and we 
introduced the average equation of state for the condensate. Following the well-known result for scalar fields 
oscillating in the potential Eq.(\ref{Eq:potmin}), we have $w_B = \frac{k-2}{k+2}$.
The first equation can be integrated to give \cite{Clery:2021bwz}
\be
\label{Eq.rhoB}
\rho_B = \rho_{\rm end}\left(\frac{\aend}{a}\right)^{\frac{6k}{k+2}}
\ee
where $\rend = 3M_P^2H_{\rm end}^2$ is the energy density in the entire background at the end of inflation. 
Using this solution to express the condensate amplitude during the oscillations, $|B|=M_P \left(\frac{\rho_B}{3\lambda 
M_P^4}\right)^{\frac{1}{k}}M_P$, we can integrate the second equation and obtain the evolution of radiation energy 
density in the form of Higgs bosons as a function of the scale factor for $\aend<a<\arh$
\bea
\label{rhoR}
\rho_R(a) &\simeq & \frac{3^{\frac{1-k}{k}}N_h M_P^4\lambda^{1/k} k^5}{2560\pi }\left(80\xi^2 -120\xi 
+49\right)\left(\frac{\pi}{2}\right)^{5/2}\left(\frac{\Gamma(\frac{1}{2}+\frac{1}{k})}{\Gamma(\frac{1}{k})}\right)^5 
\nonumber\\
&&\times \left(\frac{k+2}{8k-14}\right)\sum\limits_{n=1}^{\infty}n^5|\mathcal{P}^{(2)}_n|^2\left(\frac{\rend}{M_P^4}\right)^{\frac{2k-1}{k}}\left(\frac{\aend}{a}\right)^4
\eea
We introduce the same notation as \cite{Co:2022bgh, Barman:2022qgt}
\be
\rho_R(a) \simeq \alpha_k^\xi  M_P^4 \left(\frac{k+2}{8k-14}\right)\left(\frac{\rend}{M_P^4}\right)^{\frac{2k-1}{k}}\left(\frac{\aend}{a}\right)^4
\ee
with 
\be
 \alpha_k^\xi = 3^{\frac{1-k}{k}}\frac{N_h M_P^4\lambda^{1/k} k^5}{2560\pi}\left(80\xi^2 -120\xi +49\right)\left(\frac{\pi}{2}\right)^{5/2}\left(\frac{\Gamma(\frac{1}{2}+\frac{1}{k})}{\Gamma(\frac{1}{k})}\right)^5  \sum\limits_{n=1}^{\infty}n^5|\mathcal{P}^{(2)}_n|^2
\ee

From this expression of the radiation energy density, defining the end of reheating as $\rho_B(\arh) = \rho_R(\arh)$, we obtain the expression of the reheating temperature for this process \cite{Co:2022bgh, Barman:2022qgt}
\be
\frac{\pi^2 g^*_{\rm RH}}{30} \Trh^4 = M_P^4 \left(\frac{\rend}{M_P^4}\right)^{\frac{4k-7}{k-4}}\left(\frac{\alpha_k^\xi (k+2)}{8k-14}\right)^{\frac{3k}{k-4}}
\label{grav_TRH}
\ee
\begin{figure}[htb!]
    \centering
\includegraphics[width=0.6\linewidth]{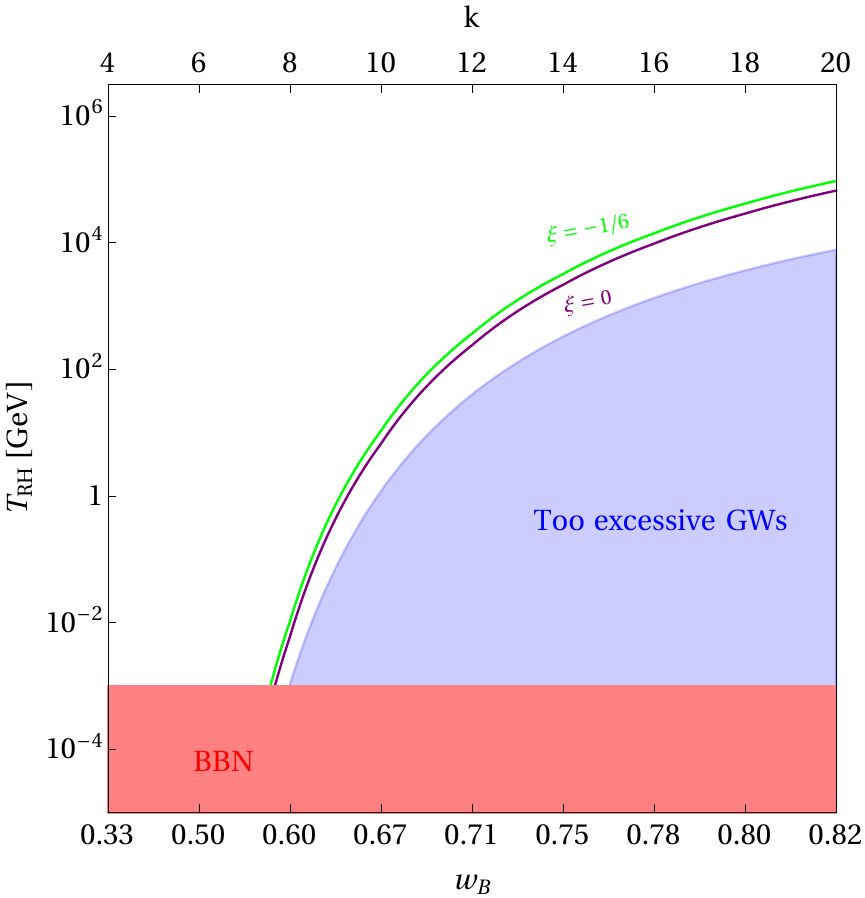}
    \caption{\it Variation of $\, \Trh$ as a function of the equation of state during the reheating era, $w_B$, for $\xi=-\frac{1}{6}$ (green) and $\xi=0$ (purple). The red-shaded region is excluded by BBN, while the blue region is excluded by a too-excessive enhancement of the primordial GWs spectrum.}
    \label{fig:reheating_nonmin}
\end{figure}
We show in Figure (\ref{fig:reheating_nonmin}) the evolution of $\Trh$ as a function of the shape of the potential near the minimum, $k\in[6,20]$, 
requiring that $\xi=-\frac{1}{6}$ to have successful inflation from these background vector fields. In this plot we considered the values $N_h=4$, $\rho_{\rm end} = (5\times 10^{15})^4 \, \gev^4$ and $\lambda$ introduce in the first section Eq.(\ref{Eq.potnormalization}). As we can see, the needed non-minimal coupling to the gravity of the vector fields for successful inflation imposes an unavoidable lower bound on the reheating temperature reached through the perturbative processes depicted in this section. Hence, in the vector inflation framework, depending on 
the equation of state during reheating, i.e., depending on the leading order term in the expansion of the inflaton potential near the 
minimum, the reheating temperature of the Universe is constrained to be higher than the one given by the green line in Fig.(\ref{fig:reheating_nonmin}).

We also show the region (blue-shaded) 
in the parameter space $(\Trh, w_B)$, which is currently excluded by too excessive enhancement of primordial gravitational waves 
(GWs). Indeed, GWs generated by quantum fluctuations during 
inflation, followed by a reheating era where the inflaton energy redshifts faster than radiation, results in an enhancement 
of GWs spectrum \cite{Barman:2022qgt,Chakraborty:2023ocr,Barman:2023ktz}. 
This effect places a constraint from excessive GWs as dark radiation at BBN and CMB time \cite{Figueroa:2018twl, Yeh:2022heq} and offers a signal with a distinctive spectrum depending on the equation of state during $w_B$. An important result of this analysis is that gravitational reheating is possible after vector inflation as opposed to the standard scalar case, where the GWs constraints already exclude the most simple gravitational reheating scenario \cite{Barman:2022qgt}.  In addition to the graviton exchange process, the necessary non-minimal coupling to gravity enhanced the reheating temperature of gravitational vector reheating and this scenario can be probed by future GWs detectors such as DECIGO \cite{Seto:2001qf,Kawamura:2006up,Yagi:2011wg} and BBO \cite{Crowder:2005nr,Corbin:2005ny,Harry:2006fi}. We now look for more efficient reheating mechanisms, also naturally present in the context of vector inflation.

%%%%%%%%%%%%%%%%%%%%%%%
\subsection{Decay of the inflaton background towards fermions}

Besides the unavoidable couplings to the gravitational sector required to achieve successful inflation,  the 
additional vector fields 
can also couple to fermions of the SM through gauge couplings. Let's consider a new $U(1)_X$ gauge symmetry 
\cite{Langacker:1984dc, Leike:1998wr, Langacker:2008yv} associated 
with the vector fields responsible for inflation and ask SM fermions $f$
to be charged under this new gauge group with charges $q_f$ 
and gauge coupling $g_X$.
We can, of course, image much larger groups of symmetry for this additional "dark" gauge. However, usually, these 
extensions would 
embed the SM gauge group in a larger set of non-abelian symmetry transformations, many of which mix quarks and 
leptons. We consider 
the simple possibility of only one additional abelian gauge group $U(1)_X$. 
This additional gauge coupling allows for the vector inflatons to decay 
into SM fermions at tree level, which can further decay to SM states. We emphasize that the vector inflaton scenario, contrary to the standard scalar
inflaton, provides a natural large coupling to fermionic states. 
Indeed, in the vanilla reheating scenario for a scalar inflaton $\phi$, 
we rely on an effective Yukawa-like coupling of the form $\propto y\phi \bar{f}f$, with SM states. However, this coupling should only emerge as an effective coupling generated by heavy fermions integrated out and is not a fundamental coupling of the theory, $\phi$ is not charged under the SM gauge group. 
On the other hand, in the vector inflaton scenario, gauge coupling to fermions arising from a GUT framework is natural and only constrained by the fundamental symmetry imposed on the Universe. \\

To compute the reheating temperature generated by the decay of the inflaton into fermions, we consider the following model for interaction between SM fermions and vector inflaton fields 
\be
\mathcal{L}\supset -\frac{1}{4} F_{\mu\nu}F^{\mu\nu} -q_fg_X\bar{f}A^\mu\gamma_\mu f + i\bar{f}\slashed{\partial}f
\ee
\bea
\scalebox{1}{
\begin{tikzpicture}[baseline={-0.1cm}]
  \begin{feynman}[every blob={/tikz/fill=gray!30,/tikz/inner sep=2pt}]
    \vertex (i1) at (1, 1.2) {\(\bar f\)};
    \vertex (i2) at (1,-1.2) {\(f\)};
    \vertex (m) at (0,0);
    \vertex (e1) at (-1.5,0) {\(A_\m\)} ;
    \diagram* {
      (m) -- [boson, style=red] (e1),
      (i1) -- [fermion] (m),
      (m) -- [fermion] (i2)}; \end{feynman}
\end{tikzpicture}
\label{Fig:A->ff}
}
\eea
The fermions can be considered massless in comparison to the energy transferred
through the inflaton decay.
The squared amplitude associated with the process depicted above is then, for the contribution of one of the three vector fields labeled $(a)$, given by 
\be
|\bar{\mathcal{M}}^{(a)}_n|^2 = 2q_f^2g_X^2|B|^2|\mathcal{P}_n|^2\left(E_n^2-4(\epsilon_i^{(a)}\,p_i)^2\right)
\ee
where the subscript $n$ corresponds to the {\it n-th} oscillating mode of each background vector field $B(t)$. We can compute from this transition amplitude the rate of production of these fermions from the background
\be
R=\frac{2\times 3}{6\pi} q_f^2 g_X^2|B|^2 \sum\limits_{n=1}^{+\infty} |\mathcal{P}_n|^2E_n^2
\ee
the factor of 2 stands for the fact that two fermions are produced per decay, whereas the factor 3 corresponds to the three orthogonal vector fields. Note that 
We now have to solve the system of Boltzmann equations (\ref{Eq:boltzmann})
for radiation (in the form of fermionic particles) and background energy densities. Solutions for $\rho_R$ are the following in the limit $a\gg a_{\rm end}$ 
\bea
    \rho_R(a) &&\simeq \sum\limits_{i} (q_f^i)^2 N_f^i \times \frac{3^{\frac{1-k}{k}}g_X^2 \lambda^{1/k}M_P^4k^3}{\pi} \left(\frac{\pi}{2}\right)^{3/2} \left(\frac{\Gamma(\frac{1}{k} + \frac{1}{2})}{\Gamma(\frac{1}{k})}\right)^3 \sum\limits_{n=1}^{\infty}n^3|\mathcal{P}_n|^2 \nonumber \\
    && \times \left(\frac{k+2}{14-2k}\right)\left(\frac{\rend}{M_P^4}\right)^{\frac{k-1}{k}}\left(\frac{\aend}{a}\right)^{\frac{6k-6}{k+2}}~~~~~~~~ (k<7) \\
    \rho_R(a) && = \sum\limits_{i} (q_f^i)^2 N_f^i \times \frac{343 g_X^2\lambda^{1/7}M_P^4}{3^{6/7}\pi} \left(\frac{\pi}{2}\right)^{3/2} \left(\frac{\Gamma(\frac{9}{14})}{\Gamma(\frac{1}{7})}\right)^3 \sum\limits_{n=1}^{\infty}n^3|\mathcal{P}_n|^2 \nonumber \\ 
    && \times \left(\frac{\rend}{M_P^4}\right)^{\frac{6}{7}}\left(\frac{\aend}{a}\right)^4\log{\left(\frac{a}{\aend}\right)}~~~~~~~~~~~~~~~~(k=7)\\
    \rho_R(a) && \simeq \sum\limits_{i} (q_f^i)^2 N_f^i \times \frac{3^{\frac{1-k}{k}}g_X^2 \lambda^{1/k}M_P^4k^3}{\pi} \left(\frac{\pi}{2}\right)^{3/2} \left(\frac{\Gamma(\frac{1}{k} + \frac{1}{2})}{\Gamma(\frac{1}{k})}\right)^3 \sum\limits_{n=1}^{\infty}n^3|\mathcal{P}_n|^2 \nonumber \\
    && \times \left(\frac{k+2}{2k-14}\right)\left(\frac{\rend}{M_P^4}\right)^{\frac{k-1}{k}}\left(\frac{\aend}{a}\right)^4~~~~~~~~~~~~~(k>7)
\eea
where $N_f^i$ accounts for the number of fermionic states for each SM fermion family produced by the vector inflaton decays. The assignment of charge under $U(1)_X$ of each fermion is given in the following section, in table \ref{table:2}. To evaluate the reheating temperature, we take for simplicity the same charge for every fermion under $U(1)_X$. 
From these expressions, we can compute the associated reheating temperature as a function of the model parameters for each value of $k$
\bea
    &&  \frac{\pi^2 g^*_{\rm RH}}{30} \Trh^4 = M_P^4 \left(\frac{\alpha_k (k+2)}{14-2k}\right)^k ~~~~~~~~~~~~~~~~~~~~~~~~~~ (k<7)\\
    \nonumber
    && \frac{\pi^2 g^*_{\rm RH}}{30} \Trh^4 \sim M_P^4 (\alpha_7)^7  ~~~~~~~~~~~~~~~~~~~~~~~~~~~~~~~~~~~~~~~ (k=7) \\
    \nonumber 
    && \frac{\pi^2 g^*_{\rm RH}}{30} \Trh^4 = M_P^4 \left(\frac{\rho_{\rm end}}{M_P^4}\right)^{\frac{k-7}{k-4}}\left(\frac{\alpha_k (k+2)}{2k-14}\right)^{\frac{3k}{k-4}} ~~~~~~~~ (k>7) 
\eea
with
\be
\alpha_k = \sum\limits_{i} (q_f^i)^2 N_f^i \frac{3^{\frac{1-k}{k}}g_X^2 \lambda^{1/k}k^3}{\pi} \left(\frac{\pi}{2}\right)^{3/2} \left(\frac{\Gamma(\frac{1}{k} + \frac{1}{2})}{\Gamma(\frac{1}{k})}\right)^3 \sum\limits_{n=1}^{\infty}n^3|\mathcal{P}_n|^2
\ee
where, in the specific case of $k=7$, we neglected the logarithmic dependence of the radiation energy density with respect to the scale factor to obtain an approximate reheating temperature. 
In our numerical analysis, we underline that the true numerically evaluated value is used for $k=7$.
We also note that for $k \leq 7$, inflaton decay provides a reheating temperature 
independent of $\rend$, which is a usual result for the generic 
decay rate of inflaton towards bosons or fermions. 
On the other hand, 
for $k>7$, the reheating temperature depends strongly on $\rend$. We provide in the table \ref{table:1} the sums of Fourier coefficients, numerically evaluated, needed to compute the reheating temperature from the decay of vector inflaton towards SM fermions.

\begin{table}[htbp!]
    \centering
    \begin{tabular}{|c|c|c|}
    \hline
        $k$  & $\sum\limits_{n=1}^{\infty} n^3|\mathcal{P}_n|^2$ & $\sum\limits_{n=1}^{\infty} n^2|\mathcal{P}_n|^2$ \\
        \hline
        2 &  0.250 & 0.250 \\
        \hline
        4 &   0.241  & 0.232  \\
        \hline
        6 &    0.244  &  0.224 \\
        \hline
        8 &    0.250  & 0.219 \\
        \hline
        10 &     0.257  & 0.217 \\
        \hline
        12 &  0.264  & 0.214 \\
        \hline 
        14 &    0.270 & 0.213 \\
        \hline
        16  &     0.276 &  0.212 \\
        \hline
        18 &      0.281  & 0.211\\
        \hline
        20 &      0.287  &  0.210 \\
        \hline
        
    \end{tabular}
    \caption{Sums of Fourier coefficients appearing in the decay rates of the vector inflaton background towards fermions. We chose some values of the equation of state parameter $k$ during reheating.}
    \label{table:1}
\end{table}

\begin{figure}[htb!]
    \centering
\includegraphics[width=0.5\linewidth]{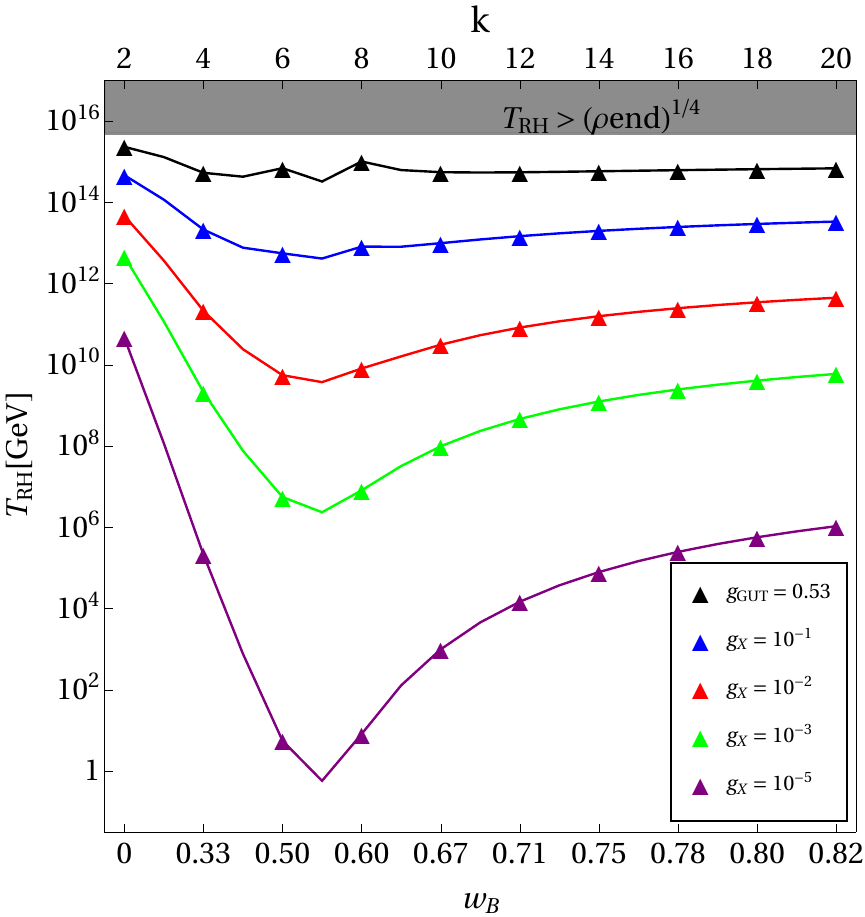}
    \caption{\it Variation of $\Trh$ as a function the 
    equation of state during reheating $w_B$, for 
    different values of the coupling between massless fermions and background fields $g_X$. The triangles show the physical choices of even values of $k$ for vector inflaton potential. The assignment of charge under $U(1)_X$ of each fermion is given in table \ref{table:2}, where we consider $x=1$.}
\label{fig:reheating_decay_fermions}
\end{figure}
We illustrate in Fig.(\ref{fig:reheating_decay_fermions}) the reheating temperature associated 
with the decay of the vector inflaton background as a function of the equation of state of the inflaton fluid for different 
values of the gauge coupling $g_X$. However, the charges $q_f$ 
are not completely free. Indeed,
asking for the Standard Model terms to be invariant under $U(1)_X$
limits the charges $q_f$ allowed. We show in table \ref{table:2}
the assignment of charges for fermions as a function of a free parameter $x$ \cite{Appelquist:2002mw, Okada:2016tci}.
To illustrate the result of the vector inflaton decay to reheat the Universe, we considered in Fig.(\ref{fig:reheating_decay_fermions}) $x=1$. We 
underline that the $U(1)_X$ gauge lets the freedom 
to choose arbitrary large values of $x$, leading to 
potentially very large couplings of fermions to 
vector inflaton. In the specific case of $x=0$, we recover the well known $B-L$ gauge \cite{Davidson:1978pm,Mohapatra:1980qe, Marshak:1979fm, Wetterich:1981bx, Masiero:1982fi, PhysRevD.27.254, Buchmuller:1991ce}.
\begin{table}[htbp!]
    \centering
    \begin{tabular}{|c|ccc|c|}
    \hline
          & $SU(3)_c $ & $SU(2)_L$ & $U(1)_Y$ & $U(1)_X$  \\
        \hline
        $q_L^i$ &  {\bf 3} & {\bf 2} & 1/6 & $(1/6)x +1/3$\\
        \hline
        $u_R^i$ &  {\bf 3} & {\bf 1} & 2/3 & $(2/3)x +1/3$\\
        \hline
        $d_R^i$ &  {\bf 3} & {\bf 1} & -1/3 & $(-1/3)x + 1/3$\\
        \hline
        $L_L^i$ &  {\bf 1} & {\bf 2} & -1/2 & $(-1/2)x -1$\\
        \hline
        $e_R^i$ &  {\bf 1} & {\bf 1} & -1 & $-x -1$\\
        \hline
        $N_R^k$ &  {\bf 1} & {\bf 1} & 0 & $-1$\\
        \hline 
        {\bf H} &  {\bf 1} & {\bf 2} & 1/2 & $(1/2)x$\\
        \hline
         S &  {\bf 1} & {\bf 1} & 0 & $+2$ \\
        \hline
    \end{tabular}
    \caption{Fermions and scalars gauge charges \cite{Appelquist:2002mw, Okada:2016tci}. In addition to the SM content, the RHN $N_R^i$, $(i = 1, 2, 3)$ and a complex scalar $S$ are introduced. The $U(1)_X$ charges of fields are determined by the parameter $x$. If $x$ is set to zero, we recover the $U(1)_{B-L}$ gauge group \cite{Davidson:1978pm,Mohapatra:1980qe, Marshak:1979fm, Wetterich:1981bx, Masiero:1982fi, PhysRevD.27.254, Buchmuller:1991ce}.}
    \label{table:2}
\end{table}
As it can be seen, the Universe reaches a high reheating temperature up to $T_{\rm RH}\sim 10^{13} \, \gev$, for $w>0.6$ ($k>8$), and $g_X \sim 0.1$.
This is not surprising as the gauge coupling at high
scale in a unified scenario is usually quite high. Moreover, 
there are no kinetic suppressions due to effective mass, 
which are usually present with Yukawa-type couplings
to the inflaton field \cite{Garcia:2021iag, Garcia:2020wiy}.
This value of gauge coupling is, moreover, natural from a GUT point of view, where
$g_X\simeq g_{GUT}\simeq 0.5$. This corresponds to an almost instantaneous perturbative reheating process after inflation. 
If one wants to be more precise, taking into account the dependence of the gauge coupling with the energy, one should consider the amplitude 
\beq
|\bar{\mathcal{M}}^{(a)}_n|^2 = 2q_f^2|B|^2
\times g_X^2(E_n)|\mathcal{P}_n|^2\left(E_n^2 -4(\epsilon_i^{(a)}\,p_i )\right)
\eeq
There is no analytical solution for $\rho_R$ in this case,
and $g_X(E_n)$ depends on the breaking scheme of
the unified group. As an example, we show in Fig.(\ref{fig:reheating_decay_fermions}), black 
line, 
the case of the breaking pattern $SO(10) \rightarrow SU(4)\times SU(2)_L\times U(1)_R ~[{\bf 
126}] \rightarrow SU(3)_c\times SU(2)_L\times U(1)_Y$ 
where $g_X(T_{\rm GUT}) = 0.53$, with $T_{\rm GUT} = 1.7\times 10^{15} \, \gev$. 
The gauge coupling runs very slightly down to its value at $E_n=M(\arh)$. One of the main conclusion 
of our analysis, is that for GUT--like couplings, 
we expect a large reheating temperature through the efficient decay of the vector inflaton into charged fermions. 
We underline that this gauge coupling of the vector inflaton fields cannot induce spontaneous mixing of their polarization while evolving in the vacuum, as it would violate the conservation of angular momentum. Hence, we expect the orthogonality condition of inflaton vector field polarizations to be conserved.

%%%%%%%%%%%%%%%%%%%%%%%
\subsection{Effective couplings of inflaton background}

In the preceding section, we considered that the vector fields could decay through their 
couplings to massless SM fermions. 
However, we can imagine a spectrum where only heavy fermions are
coupled to the background vector fields. In this case, if some of the heavy fermions
are heavier than the (effective) mass of the background fields,
they must be integrated out. Under a
spontaneous symmetry-breaking mechanism that can also provide
masses to vector fields and fermions via the VEV of a scalar
field (Stueckelberg and Higgs mechanism), some heavy fermions can be heavier than the massive vector fields. These heavy fermions can also be simultaneously
charged under other gauges, such as SM. In this case, there is an inevitable effective kinetic mixing generated between the background vector and SM gauge fields \cite{Holdom:1985ag}, associated with the gauge $U(1)_X \times U(1)_Y$,
\beq
\mathcal{L}\supset -\frac{1}{4} F_{\mu\nu}F^{\mu\nu}  -\frac{1}{4} G_{\mu\nu}G^{\mu\nu}  -\frac{\chi}{2} F_{\mu\nu}G^{\mu\nu} +\frac{1}{2}M^2 A_\mu A^\mu 
\eeq
where $G$ is 
the field strength of any SM hypercharge gauge field and $F$ is the field strength of background vector 
field $A_\mu$. From heavy fermions at a mass scale $m_j > M$ circulating in the loop, we expect a kinetic mixing of the form \cite{Cheung:2009qd}
\beq
\chi  = \frac{g_{\rm SM}g_X}{16 \pi^2} \sum\limits_{j} q_{\rm SM}^j q_X^j \log{\left(\frac{m_j^2}{\mu^2}\right)}
\eeq
where $\mu$ is a renormalization scale, that we can consider to be close to the background vector field mass scale, $M$, that defines the momentum transfer of the processes taking place during reheating. Hence, this kinetic mixing can be suppressed if heavy fermions have the same masses and opposite charges. If this is not the case, we typically expect a mixing of the order $\chi\simeq 10^{-3}~g_X$ where we considered the logarithm factor to be $\sim \mathcal{O}(1)$. 
\\

We perform a rotation in the field space that brings to a basis where fields have canonical kinetic terms and that preserves the $A_\mu$ fields as mass eigenstates of mass $M$ \footnote{We underline that such a field redefinition does not spoil the orthogonality of the polarization among the background vector fields, as they remain unchanged under the appropriate rotation.}.  After field redefinition, we can see that the effective kinetic mixing has induced an effective charge of SM fermions under $U(1)_X$, as well as a small shift in the effective charge of invisible fermions under $U(1)_X$, with the effective charge given by 
\beq
\tilde{q}_X \simeq q_X -\chi \frac{g_{\rm SM}}{g_X}q_{\rm SM}
\eeq
where $q_X$ and $q_{\rm SM}$ are the bare charges of the fermion considered under $U(1)_X$ and $U(1)_Y$ respectively. Hence, the effect of kinetic mixing is to allow for a reheating from vector fields decay towards SM fermions even without direct couplings between the vector fields and SM fermions. The associated amplitude is suppressed by $\sim  10^{-4}$ in comparison with the case of direct couplings of SM with $U(1)_X$ gauge. The results of Fig.(\ref{fig:reheating_decay_fermions}) can be adapted with this suppression factor. We expect a reheating temperature $\Trh \sim 10^{11} \rm~GeV$ for equation of states $w_B = 0$, and $g_X = g_{\rm GUT} = 0.53$, $g_{\rm SM} = 0.1$, while for $w_B \gg 0.6$, we expect $\Trh \sim 10^{7}\rm~GeV$.  
\\

Besides the kinetic mixing, heavy fermions running in the loop will also generate effective
anomalous couplings between the background fields and the SM
gauge fields through generalized Chern Simons terms (GCS) and
axionic couplings. We underline that such anomalous
effective couplings are generic \cite{DHoker:1984izu, DHoker:1984mif, Preskill:1990fr, 
Anastasopoulos:2006bs, Anastasopoulos:2006cz,
Anastasopoulos:2007qm, Anastasopoulos:2007fgu,
Anastasopoulos:2008jt, 
Dedes:2012me, Ismail:2017ulg, Arcadi:2017jqd, Dror:2017nsg, Michaels:2020fzj, Craig:2019zkf, Anastasopoulos:2022ywj} and can arise from a complete
anomaly-free theory, considering the SM spectrum and additional
massless fermions, vector fields, and scalars.  They take the form \cite{Kiritsis:2002aj, Anastasopoulos:2005ba,
Anastasopoulos:2006bs, Anastasopoulos:2006cz,
Anastasopoulos:2007qm, Anastasopoulos:2008jt, Anastasopoulos:2022ywj} 
\bea
&&\Gamma^{AYY}_{\mu\nu\rho\sigma}|_{\rm axion} ~\propto g_Xg_jg_k t_{Ajk} \frac{p_\mu}{p^2}\levi_{\nu\rho\sigma\tau}k_2^\sigma k_1^\tau\\
&&\Gamma^{AYY}_{\mu\nu\rho\sigma}|_{\rm GCS} ~~\propto g_Xg_jg_k t_{Ajk}\levi_{\mu\nu\rho\sigma}\left( k^\sigma_2 - k^\sigma_1  \right)\\
&&\Gamma^{AAY}_{\mu\nu\rho\sigma}|_{\rm axion} ~\propto g_X^2g_i t_{AAi} \left(\frac{k_{1\nu}}{k_1^2}\levi_{\rho\mu\tau\sigma} + \frac{k_{2\rho}}{k_2^2}\levi_{\mu\nu\tau\sigma}\right)k_2^\sigma k_1^\tau\\
&&\Gamma^{AAY}_{\mu\nu\rho\sigma}|_{\rm GCS} ~~\propto g_X^2g_i t_{Ajk}\levi_{\mu\nu\rho\sigma}\left( p^\sigma_2 - p^\sigma_1  \right)
\eea
where $(k_1, k_2)$ are the outgoing momenta and $(p, p_1,p_2)$ the ingoing momenta. The different effective vertices are depicted in the figure below.
\bea
&&
\scalebox{0.85}{
\begin{tikzpicture}[baseline={-0.1cm}]
  \begin{feynman}[every blob={/tikz/fill=gray!30,/tikz/inner sep=2pt}]
    \vertex (i1) at (0.75, 1.2) {\(Y_\n\)};
    \vertex (i2) at (0.75,-1.2) {\(Y_\r\)};
    \vertex (f1) at (0,0.75);
    \vertex (f2) at (0,-0.75);
    \vertex (f3) at (-1.25,0);
    \vertex (e1) at (-2.5,0) {\(A_\m\)} ;
    \diagram* {
      (f1) -- [boson, style=red] (i1),
      (f1) -- [fermion] (f2),
      (i2) -- [boson, style=red] (f2),
      (f3) -- [fermion, edge label=\(\textrm{heavy $\psi$}\)] (f1),
      (f2) -- [fermion] (f3),
      (e1) -- [boson, style=red] (f3) }; \end{feynman}
\end{tikzpicture}}
\to
\scalebox{0.85}{
\begin{tikzpicture}[baseline={-0.1cm}]
  \begin{feynman}[every blob={/tikz/fill=gray!30,/tikz/inner sep=2pt}]
    \vertex (i1) at (0.75, 1.2) {\(Y_\n\)};
    \vertex (i2) at (0.75,-1.2) {\(Y_\r\)};
    \vertex (f3) at (-0.25,0);
    \vertex (e1) at (-1.5,0) {\(A_\m\)} ;
    \diagram* {
      (f3) -- [boson, style=red] (i1),
      (i2) -- [boson, style=red] (f3),
      (e1) -- [boson, style=red] (f3) }; \end{feynman}
\end{tikzpicture}}
~,
\scalebox{0.85}{
\begin{tikzpicture}[baseline={-0.1cm}]
  \begin{feynman}[every blob={/tikz/fill=gray!30,/tikz/inner sep=2pt}]
    \vertex (i1) at (0.75, 1.2) {\(Y_\n\)};
    \vertex (i2) at (0.75,-1.2) {\(Y_\r\)};
    \vertex (f1) at (0,0.5);
    \vertex (f2) at (0,-0.5);
    \vertex (ff3) at (-0.5,0);
    \vertex (f3) at (-1.25,0);
    \vertex (e1) at (-2.5,0) {\(A_\m\)} ;
    \diagram* {
      (f1) -- [boson, style=red] (i1),
      (f1) -- [fermion] (f2),
      (i2) -- [boson, style=red] (f2),
      (ff3) -- [fermion, edge label=\(\textrm{heavy $\psi$}\)] (f1),
      (f2) -- [fermion] (ff3),
      (ff3) -- [scalar] (f3),      
      (e1) -- [boson, style=red] (f3) }; \end{feynman}
\end{tikzpicture}}
\to
\scalebox{0.85}{
\begin{tikzpicture}[baseline={-0.1cm}]
  \begin{feynman}[every blob={/tikz/fill=gray!30,/tikz/inner sep=2pt}]
    \vertex (i1) at (0.75, 1.2) {\(Y_\n\)};
    \vertex (i2) at (0.75,-1.2) {\(Y_\r\)};
    \vertex (ff3) at (-0.25,0);
    \vertex (f3) at (-1,0);
    \vertex (e1) at (-2,0) {\(A_\m\)} ;
    \diagram* {
      (ff3) -- [boson, style=red] (i1),
      (i2) -- [boson, style=red] (ff3),
      (ff3) -- [scalar] (f3),
      (e1) -- [boson, style=red] (f3) }; \end{feynman}
\end{tikzpicture}}
\label{Fig:A->YY}\\
&&\scalebox{0.85}{
\begin{tikzpicture}[baseline={-0.1cm}]
  \begin{feynman}[every blob={/tikz/fill=gray!30,/tikz/inner sep=2pt}]
    \vertex (i1) at (-0.75, 1.2) {\(A_\m\)};
    \vertex (i2) at (-0.75,-1.2) {\(A_\n\)};
    \vertex (f1) at (0,0.75);
    \vertex (f2) at (0,-0.75);
    \vertex (f3) at (1.25,0);
    \vertex (e1) at (2.5,0) {\(Y_\r\)} ;
    \diagram* {
      (f1) -- [boson, style=red] (i1),
      (f1) -- [fermion] (f2),
      (i2) -- [boson, style=red] (f2),
      (f3) -- [fermion, edge label'=\(\textrm{heavy $\psi$}\)] (f1),
      (f2) -- [fermion] (f3),
      (e1) -- [boson, style=red] (f3) }; \end{feynman}
\end{tikzpicture}}
\to
\scalebox{0.85}{
\begin{tikzpicture}[baseline={-0.1cm}]
  \begin{feynman}[every blob={/tikz/fill=gray!30,/tikz/inner sep=2pt}]
    \vertex (i1) at (-0.75, 1.2) {\(A_\m\)};
    \vertex (i2) at (-0.75,-1.2) {\(A_\n\)};
    \vertex (f3) at (0.25,0);
    \vertex (e1) at (1.5,0) {\(Y_\r\)} ;
    \diagram* {
      (f3) -- [boson, style=red] (i1),
      (i2) -- [boson, style=red] (f3),
      (e1) -- [boson, style=red] (f3) }; \end{feynman}
\end{tikzpicture}}~,~
\scalebox{0.85}{
\begin{tikzpicture}[baseline={-0.1cm}]
  \begin{feynman}[every blob={/tikz/fill=gray!30,/tikz/inner sep=2pt}]
    \vertex (i1) at (-0.75, 1.2) {\(A_\m\)};
    \vertex (i2) at (-0.75,-1.2) {\(A_\n\)};
    \vertex (f1) at (0,0.5);
    \vertex (f2) at (0,-0.5);
    \vertex (ff3) at (0.5,0);
    \vertex (f3) at (1.25,0);
    \vertex (e1) at (2.5,0) {\(Y_\r\)} ;
    \diagram* {
      (f1) -- [boson, style=red] (i1),
      (f1) -- [fermion] (f2),
      (i2) -- [boson, style=red] (f2),
      (ff3) -- [fermion, edge label'=\(\textrm{heavy $\psi$}\)] (f1),
      (f2) -- [fermion] (ff3),
      (ff3) -- [scalar] (f3),      
      (e1) -- [boson, style=red] (f3) }; \end{feynman}
\end{tikzpicture}}
\to
\scalebox{0.85}{
\begin{tikzpicture}[baseline={-0.1cm}]
  \begin{feynman}[every blob={/tikz/fill=gray!30,/tikz/inner sep=2pt}]
    \vertex (i1) at (-0.75, 1.2) {\(A_\m\)};
    \vertex (i2) at (-0.75,-1.2) {\(A_\n\)};
    \vertex (ff3) at (0.25,0);
    \vertex (f3) at (1,0);
    \vertex (e1) at (2,0) {\(Y_\r\)} ;
    \diagram* {
      (ff3) -- [boson, style=red] (i1),
      (i2) -- [boson, style=red] (ff3),
      (ff3) -- [scalar] (f3),
      (e1) -- [boson, style=red] (f3) }; \end{feynman}
\end{tikzpicture}}~~~~~~~~
\label{Fig:AA->Y}
\eea
However, we note first that in the case of on-shell ingoing and outgoing spin-one fields, the axionic couplings do not contribute to any background decay or scattering processes, as the contraction with external polarization vector with the different momenta vanishes. In addition, massive spin-one decay towards massless gauge bosons is forbidden by the Landau-Yang theorem. Then, the processes involving $A\rightarrow YY$ are forbidden. Let us consider the last possibility of vector field scattering towards a massless gauge field through the effective GCS coupling. We see that for background fields, coherently oscillating, we have a vanishing contribution as $p_1 = p_2$ in this case (it is as considering the rest frame of the massive vector fields). The only non-vanishing contribution from these effective couplings, arising from heavy fermions in the spectrum, seems to be "bremsstrahlung" like emission of massless gauge fields with a massive vector field $A_\mu$ in the final state. This process should participate in the emission of radiation from the background as well as in the fragmentation of the background (loss of coherence in the final state for the vector field) but is a negligible 
contribution to reheating.\\

Finally, we want to discuss the possibility of a next-order process that could contribute to 
reheating the Universe through effective couplings. We can consider a four-point one-loop 
amplitude with two background vector fields and two gauge fields of the SM through a "box" or 
"light to light" process \cite{PhysRev.83.776}.
\bea
&&\scalebox{1}{
\begin{tikzpicture}[baseline={-0.1cm}]
  \begin{feynman}[every blob={/tikz/fill=gray!30,/tikz/inner sep=2pt}]
    \vertex (i1) at (-1.75, 1.2) {\(A_\m\)};
    \vertex (i2) at (-1.75,-1.2) {\(A_\n\)};
    \vertex (l1) at (-0.5, 0.5);
    \vertex (l2) at (-0.5,-0.5);
    \vertex (r1) at (0.5, 0.5);
    \vertex (r2) at (0.5,-0.5);
    \vertex (f1) at (1.75, 1.2) {\(Y_\r\)};
    \vertex (f2) at (1.75,-1.2) {\(Y_\s\)};
    \diagram* {
      (i1) -- [boson, style=red] (l1),
      (i2) -- [boson, style=red] (l2),
      (f1) -- [boson, style=red] (r1),
      (f2) -- [boson, style=red] (r2),
      (l1) -- [fermion] (l2),
      (l2) -- [fermion] (r2),
      (r2) -- [fermion, edge label'=\(\textrm{~~heavy $\psi$}\)] (r1),
      (r1) -- [fermion] (l1)}; \end{feynman}
\end{tikzpicture}}
~~~\to ~~~
\scalebox{1}{
\begin{tikzpicture}[baseline={-0.1cm}]
  \begin{feynman}[every blob={/tikz/fill=gray!30,/tikz/inner sep=2pt}]
    \vertex (i1) at (-1.75, 1.2) {\(A_\m\)};
    \vertex (i2) at (-1.75,-1.2) {\(A_\n\)};
    \vertex (m) at (0,0);
    \vertex (f1) at (1.75, 1.2) {\(Y_\r\)};
    \vertex (f2) at (1.75,-1.2) {\(Y_\s\)};
    \diagram* {
      (i1) -- [boson, style=red] (m),
      (i2) -- [boson, style=red] (m),
      (f1) -- [boson, style=red] (m),
      (f2) -- [boson, style=red] (m)}; \end{feynman}
\end{tikzpicture}}
    \label{fig:AA->YY_box}
\eea
Considering again heavy fermions circulating in the loop, one can extract the effective 
couplings integrating out the messengers obtaining the following effective 
Lagrangian\footnote{Similar couplings are also expected with heavy bosonic messengers like in \cite{Chowdhury:2018tzw}.} 
\cite{Anastasopoulos:2018uyu}
\bea
\sqrt{-g}\mathcal{L}_{\rm eff} &\supset &-\frac{g_X^2g_{SM}^2}{\Lambda^4 90(4\pi)^2}\left[G_{\mu\nu}G^{\mu\nu}F_{\rho\sigma}F^{\rho\sigma} + \frac{7}{4}G_{\mu\nu}\tilde{G}^{\mu\nu} F_{\rho\sigma}\tilde{F}^{\rho\sigma} \right.\nonumber\\
&& ~~~~~~~~~~~~~~~~~~ \left. +\frac{7}{2}G_{\mu\nu}\tilde{G}_{\rho\sigma}F^{\mu\nu}\tilde{F}^{\rho\sigma} +2G_{\mu\nu}G_{\rho\sigma}F^{\mu\nu}F^{\rho\sigma}\right]
\eea
where $\Lambda$ is the heavy mass scale associated with the fermions circulating in the loop, $G$ is 
the field strength of any SM gauge field, $F$ is the field strength of the background vector 
field, and $\tilde{F}^{\mu\nu} = \frac{1}{2}\levi_{\mu\nu\alpha\beta}F^{\alpha \beta}$, is the 
dual field strength. These effective couplings can generate 2 to 2 processes $A_iA_i \rightarrow SM 
~SM$ that contributes to reheating as depicted in Figure (\ref{fig:AA->YY_box}). However, this contribution is suppressed by the 4-point vertices and provides amplitude of the form 
\be
-i\mathcal{M}^{(n)} \propto i\frac{g_X^2g_{SM}^2}{1440\pi^2\Lambda^4}|B|^2\mathcal{P}^{(2)}_n E_n^4
\ee
We then have the following square amplitudes for the box diagrams involving the same polarizations or different polarizations of vector fields in the initial state
\be
|{\bar{\mathcal{M}}^{(aa)}_n}|^2 = \frac{2^2}{2}\frac{g_{SM}^4g_X^4}{132710400\pi^4\Lambda^8}|B|^4 |\mathcal{P}^{(2)}_n|^2  E_n^4 \left(11664 (\epsilon_i^{(a)}\,p_i )^4 - 4968(\epsilon_i^{(a)}\,p_i )^2 E_n^2 + 1009 E_n^4\right)
\ee
\be
|{\bar{\mathcal{M}}^{(a\neq b)}_n}|^2 = \frac{2^2}{2}\frac{9\,g_{SM}^4g_X^4}{1638400\pi^4\Lambda^8}|B|^4 |\mathcal{P}^{(2)}_n|^2  E_n^4 \left(E_n^2-4(\epsilon_i^{(a)}\,p_i )^2\right) \left(E_n^2 - 4 (\epsilon_i^{(b)}\,p_i)^2\right)
\ee
leading to the following rate of SM gauge bosons production
\be
R = \frac{331N_Y g_{SM}^4g_X^4}{22118400\pi^5\Lambda^8} |B|^4 \sum\limits_{n=1}^{\infty} E_n^8|\mathcal{P}^{(2)}_n|^2
\label{eq:box_rate}
\ee
where we consider $g_{SM} = 0.1$ and $N_Y = 12$ for SM gauge fields. To compare this process of radiation production with the one of gravitational scattering 
discussed in Section \ref{sec:reheating_grav}, we compute the ratio of the anomaly-induced process rate 
(\ref{eq:box_rate}) over the gravitational rate (\ref{eq:grav_rate}), evaluated at the end of inflation
when $a=\aend$. In fact, if the anomalous induced box rate is subdominant at the end of inflation, it will stay 
subdominant during the whole reheating process, as it scales with a higher power of $E_n$ than the 
gravitational rate of radiation production. Indeed, $E_n(a)$ decreases as the Universe expands, so the ratio of the rates is always decreasing with time. 
\begin{figure}[htb!]
    \centering
\includegraphics[width=0.49\linewidth]{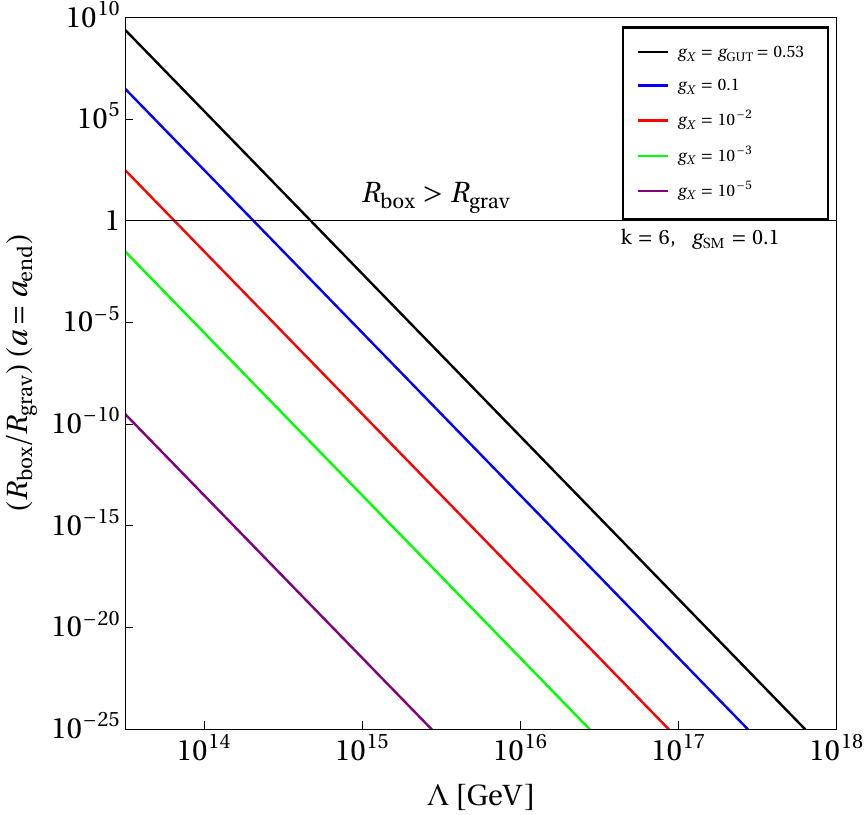}
\includegraphics[width=0.49\linewidth]{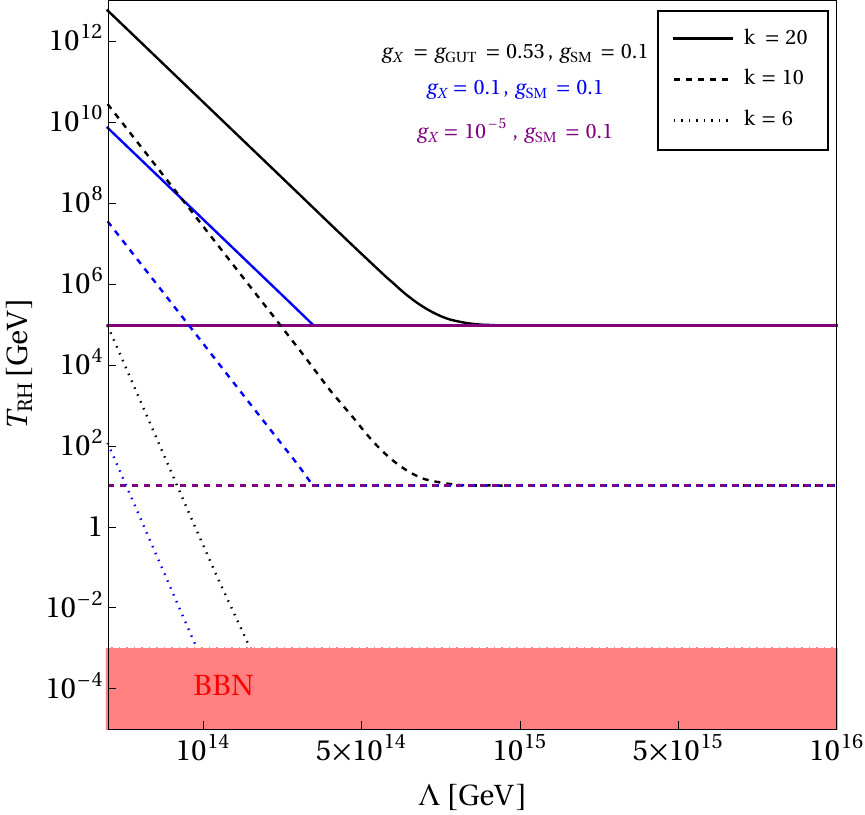}
    \caption{\it Left : Ratio between the anomalous contribution ($R_{\rm box}$) and the gravitational production ($R_{\rm grav}$) as a function
    of the effective mass scale $\Lambda$ for different values of $g_X$, at the end of inflation. Right: Reheating temperature as a function of the effective mass scale $\Lambda$, for different values of $k$ and $g_X$. Both gravitational and contributions from the anomalous effective terms are added.}
\label{fig:effective_box}
\end{figure}
We show in the left panel of Fig.(\ref{fig:effective_box}) the ratio of these rates at the end of inflation, as a function of the effective mass scale $\Lambda$, for different values of $g_X$. We see that for GUT-like gauge couplings and $\Lambda \leq 10^{15} \gev$, the effective couplings can be competitive with the gravitational effects and can be even much more efficient for a quite small effective scale $\Lambda\simeq10^{14} \gev$.
On the right panel of Fig.(\ref{fig:effective_box}), we computed
the reheating temperature obtained by combining the anomalous box
diagrams and the gravitational process. We recover the preceding 
result, noting that the reheating temperature can be efficiently 
enhanced as soon as the effective mass scale of the fermions 
ranges below $\Lambda \lesssim 10^{15}$ GeV, even reaching 
$T_{\rm RH}\gtrsim10^{12}$ GeV for GUT-type couplings $g_X$
and $\Lambda\simeq 5\times 10^{13}$ GeV.\\

In what is discussed above, we considered that fermions either from SM or at a higher energy scale, are coupled to both $U(1)_X$ and the SM gauge. However, there is a last possibility that the only fermion content coupled to the background vector fields are singlets under SM gauge. The typical example is the right-handed neutrino (RHN) that can carry a charge under $U(1)_X$ but has to be SM singlet. In the next section, we will consider the direct decay of the background vector fields towards on-shell RHN which can achieve a successful leptogenesis.
Here, we only consider the case of a very heavy RHN that has a mass $M_N>M$. We expect them to induce an effective coupling between the background vector fields, SM leptons, and Higgs. In particular, it induces radiatively a coupling between SM leptons and background vector fields of the form   
\beq
\tilde{g}_X \tilde{q}_L \sim \frac{g_X q_N y_N^2}{16\pi^2} \log\left (\frac{M_N}{\mu}\right)  
\eeq
where $q_N$ is the $U(1)_X$ charge of the RHN and $y_N$ its Yukawa coupling with SM lepton and Higgs field. The loop diagram contributing to this effect is depicted below
\bea
&&
\scalebox{1.}{
\begin{tikzpicture}[baseline={-0.1cm}]
  \begin{feynman}[every blob={/tikz/fill=gray!30,/tikz/inner sep=2pt}]
    \vertex (i1) at (0.75, 1.2) {\(L\)};
    \vertex (i2) at (0.75,-1.2)  {\(\bar{L}\)};
    \vertex (f1) at (0,0.75);
    \vertex (f2) at (0,-0.75);
    \vertex (f3) at (-1.25,0);
    \vertex (e1) at (-2.5,0) {\(A_\m\)} ;
    \diagram* {
      (f1) -- [fermion] (i1),
      (f2) -- [scalar,  edge label= \(H\)] (f1),
      (i2) -- [fermion] (f2),
      (f3) -- [fermion, edge label= \(N_R\)] (f1),
      (f2) -- [fermion, edge label= \(\bar{N}_R\)] (f3),
      (e1) -- [boson, style=red] (f3) }; \end{feynman}
\end{tikzpicture}
\label{Fig:A->LL}
}.
\eea
The production rate is suppressed by $y_N^4$ in comparison with the tree-level decays. The Yukawa coupling of the RHN is chosen to generate light neutrino masses through the see-saw mechanism, $y_N^2 \sim \frac{m_\nu M_N}{v_H^2}$, with $v_H$ the VEV of the Higgs field after electroweak symmetry breaking (we come back to the RHN Lagrangian with more detail in the next section). These radiative decays induce a reheating temperature as large as $\Trh \sim 10^{12} \rm~GeV$ for $w_B =0$ and $g_X = g_{\rm GUT} = 0.53$, $y_N \simeq 0.4 $ corresponding to $M_N = 10^{14}~\rm GeV$ and $m_\nu = 0.05~\rm eV$. For $w_B\gg 0.6$, we expect to recover $\Trh \sim 10^9\rm~GeV$ from these radiative decays, where we have adapted the results of Fig. (\ref{fig:reheating_decay_fermions}) with the suppression coming from $y_N^4$. \\
The coupling of RHN also implies a $1\rightarrow4$ decay at tree level, depicted in the diagram below 
\bea
&&
\scalebox{1.}{
\begin{tikzpicture}[baseline={-0.1cm}]
  \begin{feynman}[every blob={/tikz/fill=gray!30,/tikz/inner sep=2pt}]
    \vertex (i1) at (1., 1.) ;
    \vertex (i2) at (1.,-1.) ;
    \vertex (i3) at (1.5, 1.7) {\(\bar{L}\)} ;
     \vertex (i4) at (1.5, 0.4) {\(H\)};
    \vertex (i5) at (1.5,-0.4)  {\(H\)} ;
    \vertex (i6) at (1.5,-1.7) {\(L\)};
    \vertex (m) at (0,0);
    \vertex (e1) at (-1.5,0) {\(A_\m\)} ;
    \diagram* {
      (m) -- [boson, style=red] (e1),
      (i1) -- [fermion, edge label'=\(\textrm{~~ $\bar{N}_R$}\)] (m),
      (m) -- [fermion, edge label'=\(\textrm{~~ $N_R$}\)] (i2),
      (i3) -- [fermion] (i1),
      (i4) -- [scalar] (i1),
      (i2) -- [scalar] (i5),
      (i2) -- [fermion] (i6),}
      ; \end{feynman}
\end{tikzpicture}
\label{Fig:A->HHLL}
}
\eea
This process leads to the following rate of Higgs and SM Lepton production 
\beq
\qquad R_{\rm 4-body} \propto \frac{g_X^2q_N^2y_N^4}{\pi^5 M_N^8}|B|^2\sum\limits_{n=1}^{\infty} |\mathcal{P}_n|^2 E_n^{10}
\eeq
Thus, by evaluating the ratio of this rate for this 4-body decay and the one for the light-to-light box diagram, 
\beq
\frac{R_{\rm 4-body}}{R_{\rm box}} \simeq g_X^{-2} \left( \frac{y_N}{g_{\rm SM}}\right)^4 \left( \frac{\Lambda}{M_N}\right)^8\left(\frac{E_n}{|B|}\right)^2 
\eeq
we see that for $M_N \gtrsim \Lambda$, the 4-body decay is strongly suppressed in comparison with the box-diagram at the end of inflation, by a factor $(E_n/|B|)^2 \lesssim 10^{-10}$. It can become as efficient as the box diagram for $M_N\lesssim 10~\Lambda$, $y_N\simeq0.1$, $g_X = 0.53$, or for lower values of $g_X$. \\

Before closing this part on the reheating mechanism after vector inflation, we may have concerns about the radiative corrections at large field values induced by the couplings of the vector fields\footnote{We thank the referee for having raised such important considerations in the report.}. Indeed, the gauge coupling to fermions generates radiatively a quartic coupling for the vector inflaton fields. We can estimate from the loop that such quartic coupling is of the form
    \begin{equation}
        \sim \epsilon g_X^4 q^4 \left(A_\mu A^\mu\right)^2 
    \end{equation}
where $\epsilon$ is a loop suppression factor. Thus, if no new physics contributes to canceling this radiative correction to the potential, we expect the flatness of the potential to be spoiled at large field values and for large couplings. Supposing inflation takes place when the field values are above $\sim 10~M_P$, the flatness condition looks simply
    \begin{equation}
        g_X \lesssim 0.1 \left(\frac{\lambda}{\epsilon}\right)^{1/4}\, ,
    \end{equation}
showing that the inflaton fields coupling should stay below $g_X\lesssim 10^{-5}$ during inflation to ensure the flatness of the potential at such large field values. Interestingly, this upper bound on the gauge coupling $g_X$ coincides with the parameter space allowing for non-thermal leptogenesis to take place. It implies an upper bound on the reheating temperature in this scenario, as discussed in the next section. 

We note that this kind of constraint is not specific to the case of vector inflaton fields, as it is also present for a scalar inflaton coupled to fermions through Yukawa-like interactions. It provides, in this case, a similar constraint on the Yukawa coupling $y\lesssim 10^{-5}$, assuming a non-supersymmetric inflationary model \cite{Drees:2022aea}. However, we emphasize that we do not specify the exact UV completion of the inflationary model, and the whole spectrum of the full theory should contribute at high energies to radiative corrections. In this perspective, supersymmetry (SUSY) can render natural the fact that the CMB perturbations remain small by ensuring that radiative corrections to the inflaton potential stay under control at large field values. Thus, we may embed our phenomenological model into a supersymmetric construction, in which SUSY prevents such renormalization of the inflaton potential.  For example, in the scenario presented in
\cite{Ellis:2015kqa,Ellis:2015pla}, the construction of scalar inflaton potential is embedded within no-scale supergravity, and one could have an inflaton-matter coupling as large as $\sim O(1)$ without spoiling the inflationary predictions. We expect similar constructions to allow for vector inflatons to drive inflation. Different patterns for SUSY breaking that can accommodate successful large-field inflation are also discussed in \cite{Ellis:2015kqa}. 

In various SUSY models, however, large inflaton-matter coupling would reheat the Universe
to a very high temperature which can then lead to an overproduction of
gravitinos, even without direct couplings between gravitinos and the inflaton. The further decays of these thermally produced gravitinos could severely affect BBN and overpopulate
the Universe with DM \cite{Ellis:1983ew,Khlopov:1984pf}. An upper bound on $\Trh$ can be estimated in SUSY models \cite{Nilles:2001my}, depending on the gravitino mass scale $m_{3/2}$, $\Trh\lesssim 10^9~{\rm GeV}\left(100~{\rm GeV}/m_{3/2}\right)$. This, in turn, gives an upper bound on the inflaton-matter coupling to reheat the Universe. For $k=2$, we note that such upper bound approximately coincides with the parameter space $g_X\lesssim 10^{-5}$ for $m_{3/2}\sim 100~\rm GeV$. Yet, it is possible to consider SUSY models in which this gravitino constraint is substantially alleviated. One could require soft-SUSY masses to lie above the reheating temperature and the inflaton mass $M_{\rm SUSY}>m_\phi,\,\Trh$, preventing SUSY particles from being
produced by either thermal processes during reheating or by the decay of the inflaton. In this case, most of the sources for gravitinos production are absent. For a no-scale supergravity model, which leads to Starobinsky-like inflation, it has been argued in \cite{Dudas:2017rpa} that the constraint on reheating temperature could be then as high as $\Trh\lesssim 10^{12}~\rm GeV$. This constraint depends on the mass of the gravitino, which in this scenario, has to be $m_{3/2}\gtrsim m_\phi^2/\sqrt{3}M_P\simeq 0.1~\rm EeV$.

Still, we do not investigate further a specific UV completion of the theory, and we focus on the constraints on the effective couplings at smaller field values.

%%%%%%%%%%%%%%%%%%%%%%%%
\section{Leptogenesis}
\subsection{Right handed neutrinos}

We finally consider the possibility that the vector inflatons produce on-shell RHN, which could contribute through their decay to the generation of a lepton asymmetry. Again, we ask for the RHN also to be charged under the additional $U(1)_X$ gauge group associated with the background vector fields.
We also rely on the existence of an additional scalar degree of freedom $S$ in addition to the SM Higgs field $H$, which would be responsible for the mass scale of the new sector after a spontaneous symmetry-breaking mechanism \cite{Buchmuller:1991ce}. This additional scalar acquires a vacuum expectation value (VEV) $v_S$, different from the electroweak (EW) vacuum, $v_{H}$, and gives mass to the vector fields through the Stueckelberg mechanism. The two energy scales are decoupled, and we can assume that the additional gauge group is broken at really high energy, close to the inflationary scale. Then, the VEV of this new scalar can also generate a large Majorana mass, $M_N$, to the RHN through Yukawa-like couplings between RHN and the new scalar. Finally, the SM Higgs field can have a specific charge under the new gauge $U(1)_X$ to not spoil the minimal Yukawa sector for SM fermions. In this case, Yukawa-like interactions between RHN, SM Higgs doublet, and left-handed SM fermions are allowed and would generate Dirac mass, $m_D$ 
for neutrinos after EW symmetry breaking, depending on the VEV, $v_H$, of the SM Higgs field. 
Hence, the additional gauge, as well as the existence of RHN, can explain the tiny mass of the active neutrinos through the well-known seesaw mechanism \cite{MINKOWSKI1977421,GellMann:1980vs,Yanagida:1979as,Mohapatra:1979ia,Schechter:1980gr,Schechter:1981cv}. We remind the readers that there are three types of seesaw models, which differ by the nature of the exchanged heavy particles in the model:
\bi
\item[(i)] Type-I: SM gauge fermion singlets
\item[(ii)] Type-II: SM $SU(2)_L$ scalar triplets
\item[(iii)] Type-III: SM $SU(2)_L$ fermion triplets. 
\ei
We consider here the Type-I scenario that can be realized with only two generations of 
right-handed neutrino \cite{Frampton:2002qc,Raidal:2002xf,Ibarra:2003up,Rink:2017zrf}. 
In this model, the light active neutrinos acquire their mass through the seesaw 
suppression of the order $m_{\nu i}\sim \frac{(y_N)_{ii}^2\langle H \rangle^2}{M_{N_i}}$. 
In what follows, we will consider 
the production of one generation of RHN, but the discussion can be generalized to additional RHN families.

%%%%%%%%%%%%%%%%%%%%%%%%
\subsection{RHN production and non-thermal leptogenesis}

We consider RHN coupled to the vector fields through the coupling 
\be
\mathcal{L}_{\rm int} \supset q_N g_X\overline{N}_R\gamma_\mu A^\mu N_R \,,
\ee
where $g_X$ is the gauge coupling associated to $U(1)_X$ and $q_N$ is the charge of the 
RHN. 
This coupling allows producing the RHN out-of-equilibrium from the background, as long as they are less massive than the vector fields, i.e., $M_N<M$,\footnote{$N_R$ fields are not mass eigenstates, but $ M_N$ can approximate the mass of these fields in the limit $M_N\gg m_D$.} and the number density of RHN produced is the following at the end of the reheating phase
\be
n_N(a_{\rm RH})= \frac{\sqrt{3}q_N^2g_X^2M_P^3}{72}k^2(k+2)\left(\frac{\Gamma(\frac{1}{k}+\frac{1}{2})}{\Gamma(\frac{1}{k})}\right)^2\sum\limits_{n=1}^{\infty}n^2|\mathcal{P}_n|^2\left(\frac{\rho_{\rm RH}}{M_P^4}\right)^{1/2}\,,
\ee
where $\rRH = \frac{\pi^2g^\ast_{\rm RH}}{30} \Trh^4$. We 
also provide in the table \ref{table:1} the sums of Fourier 
coefficients, numerically evaluated, needed to compute the 
number density of RHN produced from the decay of vector inflaton.   \\

Then these RHN can also be coupled to SM leptons, $L$, and Higgs boson, $H$, through a Yukawa-like coupling, $y_N^2 \simeq m_\nu M_N/ v_H^2$, where $\langle H\rangle \equiv v_H \approx 174$ GeV is the SM Higgs doublet VEV,
\be
\mathcal{L}_{\rm int} \supset y_N \overline{L}\tilde{H}N_R
\ee
if the charge of $H$ under $U(1)_X$ is opposite to the sum of the charges of $N_R$ and $\overline{L}$. We provide the charge assignments required under SM gauge and the additional $U(1)_X$, for SM states and RHN, in table \ref{table:2}. These charge assignments ensure that SM Yukawa terms and the additional Yukawa terms for RHN are $U(1)_X$ invariant. The above coupling allows the heavy RHH to experience out-of-equilibrium decay towards Higgs bosons and leptons. The resulting CP-asymmetry is \cite{Luty:1992un, Flanz:1994yx, Covi:1996wh, Buchmuller:2005eh, Davidson:2008bu}
\be
\label{eq:cp}
\epsilon_{\Delta L} = \frac{\sum_{\alpha}[\Gamma(N_R \rightarrow L_{\alpha}+H)-\Gamma(N_R \rightarrow \overline{L}_{\alpha}+H^{*})]}{\sum_{\alpha}[\Gamma(N_R\rightarrow L_{\alpha}+H)+\Gamma(N_R \rightarrow \overline{L}_{\alpha}+H^{*})]}\, .
\ee
It is generated by the interference between tree-level and one-loop processes and can produce an out-of-equilibrium lepton asymmetry, depending on the abundance of RHN produced by the background. 
The CP asymmetry can be expressed as \cite{Buchmuller:2004nz,Kaneta:2019yjn,Co:2022bgh}
\be
\epsilon_{\Delta L}\simeq \frac{3 \delta_{\rm eff}}{16\,\pi}\,\frac{M_{N}\,m_{\nu\,,\text{max}}}{v_H^2}\,,  
\ee
where $\delta_{\text{eff}}$ is the effective CP violating phase in the neutrino mass matrix with $0\leq \delta_{\rm eff}\leq 1$,  and we take $m_{\nu,\text{max}} = 0.05$ eV as the heaviest light neutrino mass. We look for the constraints on the model of non-thermal leptogenesis \cite{Giudice:1999fb,Asaka:1999yd,Asaka:1999jb,Campbell:1992hd}. If $M_N \gtrsim
T_{\rm RH}$, the produced RHN are out of thermal equilibrium and their non-thermal decay generates a lepton asymmetry during reheating. We focus on such a parameter space where this condition is satisfied and non-thermal leptogenesis takes place. For a higher reheating temperature, we need to take into account the inverse decay and scattering processes with the thermal plasma that can be efficient as long as $T_{\rm RH} \gtrsim M_N$. This requires a detailed Boltzmann analysis, which has not been studied in this work.
The produced lepton asymmetry is eventually converted to baryon asymmetry via electroweak sphaleron processes leading to
\be
\label{eq:yb}
Y_B = \frac{n_B}{s} =  \frac{28}{79} \epsilon_{\Delta L}\,  \frac{n_{N}(\Trh)}{s} \,,
\ee
where $n_{N}(\Trh)$ is the number density of RHN at the end of reheating and $s = 2 \pi^2 g_{\rm RH} \Trh^3 /45$ is the entropy density. The final asymmetry then becomes 
\be
Y_B \simeq 8.7\times 10^{-11}\, \delta_{\rm eff} \left(\frac{m_{\nu,\text{max}}}{0.05 \,\text{eV}}\right) \left(\frac{M_{N}}{2.5\times10^{6}\,\text{GeV}}\right)\,\left.\frac{n_{N}}{s}\right|_{\Trh}\,,    
\label{Eq:baryonassym}
\ee
while the observed value, as reported by Planck~\cite{Aghanim:2018eyx}, is $Y_B^\text{obs}\simeq 8.7\times 10^{-11}$. 

\begin{figure}
    \centering
\includegraphics[width=0.48 \linewidth]{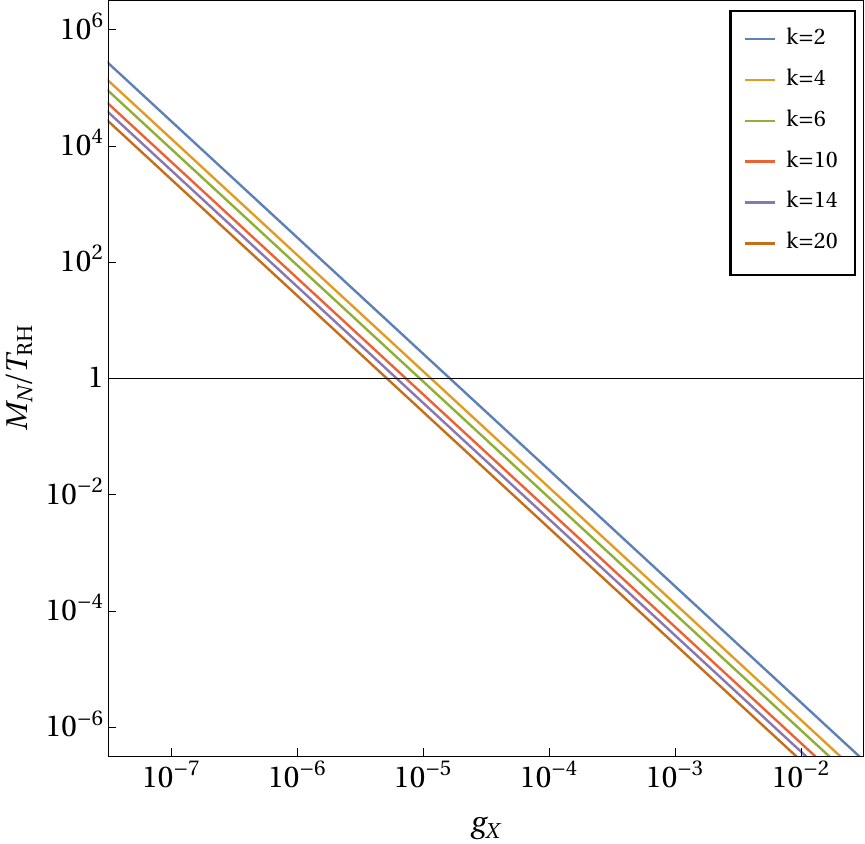}
\includegraphics[width=0.48 \linewidth]{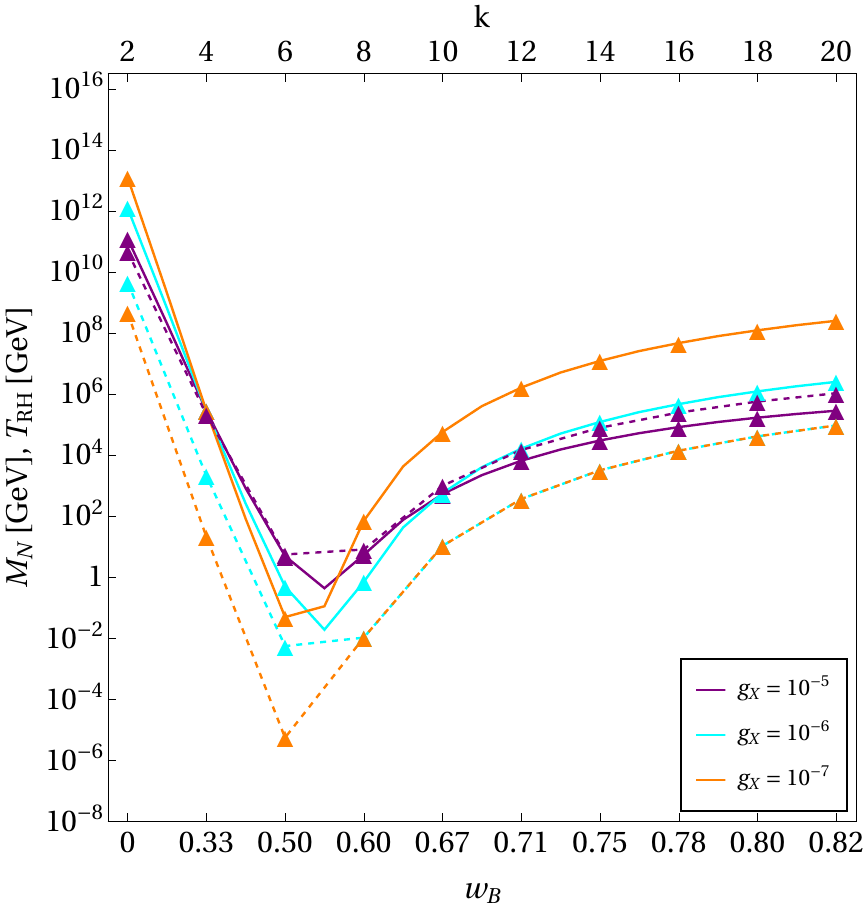}
    \caption{\it Left: Ratio $M_N/T_{\rm RH}$ obtained from Eq.(\ref{M_N_ratio}) as a function of the gauge coupling $g_X$. Different colored lines correspond to different equations of state parameter $k$. Right: RHN mass $M_N$ providing the observed BAU through non-thermal leptogenesis (solid lines) as a function of the equation of state parameter $k$, and for different values of the gauge coupling $g_X$. Reheating temperature (dashed lines) for the associated values of the gauge coupling $g_X$, as a function of the equation of state parameter $k$.}
    \label{fig:lepto_constraints}
\end{figure}

Imposing $M_N>T_{\rm RH}$ corresponds to a constraint on the magnitude of the coupling $g_X$ between vector inflatons and fermions. Indeed, it is the same coupling that drives the production of the thermal bath and the generation of the RHN abundance. Using Eq.(\ref{Eq:baryonassym}), we obtain the RHN mass $M_N$ that provides the correct baryon yield $Y_B^{\rm obs}$ for a given $T_{\rm RH}$, and we can compute the associated ratio $M_N/T_{\rm RH}$. This ratio is, in fact, independent of the reheating temperature and depends only on the coupling $g_X$ and the parameter $k$,
\begin{equation}
    \frac{M_N}{\Trh} \simeq \left(\frac{1.8\times10^{-5}}{q_Ng_X}\right)^2\frac{1}{k^2(k+2)}\left(\frac{\Gamma(\frac{1}{k})}{\Gamma(\frac{1}{k}+\frac{1}{2})}\right)^2\frac{1}{\sum\limits_{n=1}^{\infty}n^2|\mathcal{P}_n|^2}\, ,
    \label{M_N_ratio}
\end{equation}
where we take $m_{\nu,\rm max} = 0.05~\rm eV$, $\delta_{\rm eff} = 1$ and $g_{\rm RH} = 106.75$. We show the ratio $M_N/T_{\rm RH}$ as a function of the coupling $g_X$ for various values of $k$ in the left panel of Figure \ref{fig:lepto_constraints}.
We find that the reheating temperature is sufficiently small to ensure $T_{\rm RH} \lesssim M_N$, while still generating the RHN abundance for the lepton asymmetry if $g_X \lesssim 10^{-5}$. This bound is quite independent of the equation of state parameter $k$ as it can be seen in left panel of Figure \ref{fig:lepto_constraints}. For such small values of $g_X$, the reheating temperature is bounded for each value of $k$ and we find $\Trh \lesssim 5\times 10^{10}~\rm GeV$ for $k=2$, while $\Trh \lesssim 10^{6}~\rm GeV$ for $k=20$. In particular, for $g_X\lesssim 10^{-6}$ and $k > 6 $, the vector inflaton decay towards SM fermions is not efficient enough to reheat the Universe, whereas this is the gravitational portals that achieve reheating. The reheating temperature is then given by Eq.(\ref{grav_TRH}) as a function of $k$. It is determined by the value of the non-minimal coupling to gravity $\xi=-1/6$, independently of the value of $g_X$, and represents a lower bound of the reheating temperature for $k \geq 6$. However, for $k<6$, reheating can still only be achieved through the gauge coupling $g_X$. We show in the right panel of Figure \ref{fig:lepto_constraints} the reheating temperature as a function of $k$ (dashed lines) for the values of $g_X$ that allow for non-thermal leptogenesis. At large values of $k$ and small values of $g_X$, the reheating temperature is given by the gravitational portals with $\xi = -1/6$. For lower values of $k$, it is still determined by the coupling $g_X$, and especially for $k=6$ and $g_X\lesssim 10^{-7}$, it is providing a too small reheating temperature $T_{\rm RH}\lesssim\rm MeV$, excluded by BBN. 

Finally, we show in the right panel of Figure \ref{fig:lepto_constraints} the constraint on the RHN mass $M_N$ to provide the observed baryon asymmetry $Y_B^{\rm obs}$, as a function of $k$ (solid lines). For $g_X<10^{-5}$, we can see that $M_N>\Trh$, leading to non-thermal leptogenesis\footnote{The points for $k=6$ and $g_X\lesssim 10^{-6}$ are excluded by BBN because of a too low reheating temperature.}. We obtain that a RHN of mass $M_N\sim 10^{12}~\rm GeV$ can generate sufficient asymmetry for $k=2$ and $g_X=10^{-6}$, corresponding to $\Trh\sim 10^{10}~\rm GeV$, while a RHN mass $M_N\sim 10^6~\rm GeV$ is sufficient for $k=20$ and $g_X=10^{-6}$. For larger gauge coupling, the RHN mass has to be smaller so as not to overproduce the lepton asymmetry. For $M_N\ll T_{\rm RH}$, it is expected that RHN reach thermal equilibrium before their decay and leptogenesis then occur in the strong washout regime \cite{Hahn-Woernle:2008tsk,Davidson:2008bu}. In the strong washout regime, the Yukawa coupling is large enough to keep the system at equilibrium, erasing dependence on reheating temperature and the details of the production mechanism. \\

%%%%%%%%%%%%%%%%%%%%%%%%
%%%%%%%%%%%%%%%%%%%%%%%
\section{Conclusions}
\label{sec:concl}

In this work, we studied the reheating and non-thermal leptogenesis in the case of vector inflatons
$A_\mu$. We concentrated on particle perturbative 
production during the oscillating background phase, first insisting on the gravitational production induced by the presence of non-minimal coupling imposed by an isotropic and homogeneous Universe. 
We then included processes involving the graviton exchange and computed the reheating temperature by combining the minimal and non-minimal sources. 
The result illustrated in Fig.(\ref{fig:reheating_nonmin}) shows that reasonable reheating temperature can be reached only for large 
equation of state parameter $w$.
We then extended our study to decay into fermions via direct or effective couplings. 
We show that large reheating temperature can be induced by such decay due to the gauge nature of the coupling, see Fig.(\ref{fig:reheating_decay_fermions}). Furthermore, even if SM fermions are not directly coupled to the vector inflaton fields, effective charges are generated through the inevitable kinetic mixing of vector inflaton and SM $U(1)$ gauge fields.
We also study the existence of couplings appearing
through the mechanisms of anomaly cancellation. 
The presence of Chern-Simons terms allows also the reheating to proceed, dominating over the gravitational production for an effective mass scale $\Lambda \lesssim 10^{15}$ GeV as one can see in Fig.(\ref{fig:effective_box}) left. Such anomalous processes lead, however, to much lower reheating temperature than direct decays of $A_\mu$, even if it can reach $T_{\rm RH}\gtrsim 10^{12}$ GeV for a heavy fermions mass scale $\Lambda\sim 10^{13}$ GeV and GUT-type coupling $g_X$ as it can be seen in Fig.(\ref{fig:effective_box}) right. 
Finally, we looked at the possibility of generating the baryon asymmetry through a non-thermal leptogenesis process, coupling the right-handed neutrino with
$A_\mu$. Such coupling can easily be justified by gauging an extra $U(1)_X$. We show that successful non-thermal leptogenesis occurs for gauge coupling smaller than $g_X \lesssim 10^{-5}$. When the coupling is larger, thermal leptogenesis is expected to take place in the strong-washout regime, and a detailed Boltzmann analysis is required.
Finally, we noted that the nature of vectorial coupling avoids the generation of large effective mass, as is the case for a scalar inflaton, which forbids the kinematic suppression of production mechanisms. The vectorial nature of the coupling should also drastically affect the phenomenology of preheating and should be the subject of future work.

%%%%%%%%%%%%%%%%%%%%%%
%%%%%%%%%%%%%%%%
\section*{Acknowledgements}
%%%%%%%%%%%%%%%%
The authors want to thank Juan Pablo Beltran Almeida and Fredy Alexander Ochoa Perez for very fruitful discussions, as well as Marco Peloso, Mathieu Gross, Essodjolo Kpatcha, and Jong-Hyun Yoon for helping to clarify issues.  This project has received funding /support from the European Union's Horizon 2020 research and innovation program under the Marie Sklodowska-Curie grant agreement No 860881-HIDDeN, and the IN2P3 Master Projet UCMN. 
P.A. was supported by FWF Austrian Science Fund via the SAP P 36423-N.

%%%%%%%%%%%%%%%%%%%%%
\bibliographystyle{JHEP}
\bibliography{reheating}
%%%%%%%%%%%%%%%%%%

\end{document}